\documentclass[useAMS,usenatbib,a4paper]{mn2e}
\usepackage{amsmath}
\usepackage[pdftex]{graphicx}
\usepackage{epstopdf}
\usepackage[english]{babel}
\usepackage{subfigure}
\usepackage{array,multirow}
\usepackage{underscore}
\usepackage{natbib}
\usepackage{float}
\usepackage{amssymb}
\usepackage{placeins}
\usepackage{tabularx}
\usepackage{booktabs}

\newcolumntype{L}[1]{>{\raggedright\let\newline\\\arraybackslash\hspace{0pt}}m{#1}}
\newcolumntype{C}[1]{>{\centering\let\newline\\\arraybackslash\hspace{0pt}}m{#1}}
\newcolumntype{R}[1]{>{\raggedleft\let\newline\\\arraybackslash\hspace{0pt}}m{#1}}

\graphicspath{{figures/}}

\def\mnras{MNRAS}
\def\apj{ApJ}

\def\apjl{ApJL}

\def\araa{ARA\&A}
\def\pasp{PASP}
\def\nat{Nature}

\title[The dark and luminous mass structure of ETGs]{Dark matter contraction and stellar-mass-to-light ratio gradients in massive early-type galaxies}
\author[L. J. Oldham \& M. W. Auger]{Lindsay Oldham$^{1,2}$\thanks{E-mail: lindsay.oldham@cfa.harvard.edu}\thanks{Menzel Fellow} \& Matthew Auger$^2$\\
$^{1}$ Harvard-Smithsonian Center for Astrophysics, 60 Garden Street, Cambridge, MA 02138, USA \\
$^{2}$ Institute of Astronomy, University of Cambridge, Madingley Road, Cambridge CB3 0HA, UK \\
}

\pubyear{2017}

\begin{document}
\maketitle
\setcounter{page}{1}

\begin{abstract}
\noindent
We present models for the dark and luminous mass structure of 12 strong lensing early-type galaxies (ETGs). We combine pixel-based modelling of multiband HST/ACS imaging with Jeans modelling of kinematics obtained from Keck/ESI spectra to disentangle the dark and luminous contributions to the mass. Assuming a gNFW profile for the dark matter halo and a spatially constant stellar-mass-to-light ratio $\Upsilon_{\star}$ for the baryonic mass, we infer distributions for $\Upsilon_{\star}$ consistent with IMFs that are heavier than the Milky Way's (with a global mean mismatch parameter relative to a Chabrier IMF $\mu_{\alpha c} = 1.80 \pm 0.14$) and halo inner density slopes which span a large range but are generally cuspier than the dark-matter-only prediction ($\mu_{\gamma'} = 2.01_{-0.22}^{+0.19}$). We investigate possible reasons for overestimating the halo slope, including the neglect of spatially varying stellar-mas-to-light ratios and/or stellar orbital anisotropy, and find that a quarter of the systems prefer radially declining stellar-mass-to-light ratio gradients, but that the overall effect on our inference on the halo slope is small. We suggest a coherent explanation of these results in the context of inside-out galaxy growth, and that the relative importance of different baryonic processes in shaping the dark halo may depend on halo environment.

\end{abstract}


\section{Introduction}
\label{sec:chap8sec1}

Dark-matter-only simulations of a $\Lambda$CDM Universe predict that the dark matter haloes of all galaxies should look nearly self-similar, regardless of their mass scale, with a density profile which declines as $r^{-3}$ at large radii and has a central $r^{-1}$ cusp \citep{Navarro1996,Navarro2010}. However, the dominance of baryonic matter in the centres of massive ETGs may significantly modify the dark matter structure from this simple expectation, such that the halo may become either centrally contracted due to the initial infall of gas \citep{Blumenthal1986,Gnedin2004} or expanded due to dynamical heating during satellite infall \citep{ElZant2001,Nipoti2004,Laporte2012} or AGN-driven outflows \citep{Governato2012,Martizzi2012}. The halo structure is therefore key to determining the relative importance of different physical processes in ETG evolution.

However, since most gravitational probes are only sensitive to the \emph{total} galaxy mass, determining the halo structure relies on robustly decomposing the mass into its dark and luminous contributions. This problem is complicated by the fact that the form of the stellar initial mass function (IMF) in ETGs is not well understood. Whilst the diversity of environments within the Milky Way are consistent with having one `universal' IMF \citep{Bastian2010}, evidence from lensing and dynamics suggests that the IMF of massive ETGs may be heavy relative to the Milky Way \citep[e.g.][]{Auger2010b, Cappellari2012}, which stellar population modelling of gravity-sensitive spectral features has attributed to an excess of low-mass stars (i.e. to the IMF being \emph{bottom-heavy}; \citealp{vanDokkum2010a}).  Moreover, recent evidence from stellar population modelling  of optical spectra and dynamics has suggested that the IMF may also vary radially \emph{within individual galaxies}, \citep{MartinNavarro2015,vanDokkum2016,M873}, with the deviations from a Milky-Way-like IMF confined to the central regions; however, constraints from molecular gas kinematics and stellar population modelling at near-infrared wavelengths have not confirmed this \citep{Zieleniewski2016,Vaughan2017,Alton2017,Davis2016}. 

Knowledge of the IMF in massive ETGs is essential for probing halo structure; however, it is also critical for understanding ETG assembly and evolution, since the IMF is a key diagnostic of the physical conditions in which a galaxy first formed stars. Furthermore, the measurement of stellar population properties in external galaxies depends on knowledge of the IMF, such that the assumption of an incorrect IMF may lead to systematic biases in properties such as metallicity and age. Methods to simultaneously determine the dark and luminous mass structure of ETGs are therefore extremely important.

Recently, progress has been made by combining mass probes across different spatial scales such that, in the context of a well-motivated model, this dark/light degeneracy can be broken. Due to the need for multiple mass tracers, these studies have focused on massive systems in dense environments, where populations of satellite galaxies, globular clusters and planetary nebulae can be used to dynamically trace the mass out to the virial radius and so complement stellar kinematic and strong lensing constraints in the central regions. In BCGs, these studies have found haloes that are centrally less dense relative to the NFW prediction \citep{Newman2013,Oldham2016b}. On the other hand, the single study so far of group-scale ETGs has found their haloes to be mildly contracted (and consistent with the NFW model within $2\sigma$; \citealp{Newman2015}). On the scale of isolated field ellipticals, breaking the dark/light degeneracy is more difficult due to the absence of large-radius tracers and the smaller physical size of the Einstein radius, but one study of a rare double source plane lens has found evidence for strong contraction on these scales (\citealp{Sonnenfeld2012}; see also \citealp{Grillo2012}, which obtained similar constraints by combining aperture mass measurements for an ensemble of lenses assuming a fixed IMF, and \citealp{Sonnenfeld2015}, which made no IMF assumptions but weaker inference on the halo slope). These differences in the halo structure as a function of environment may represent a real trend in the relative importance of baryonic processes; however, the small number of such studies so far -- especially of isolated systems -- makes it difficult to draw meaningful conclusions. Improved techniques for extracting information from isolated strong lenses are therefore needed.

This study is a first attempt to address this need. So far, studies of the halo structure of isolated strong lenses have condensed the lensing information down to a single measure of the Einstein radius or a set of conjugate points whose positions must be focused in the source plane \citep{Grillo2012,Sonnenfeld2012,Sonnenfeld2015}; however, the extended arcs in galaxy-galaxy lenses contain much more information than this, such that a full reconstruction of the lensed images allows a much more precise inference on the halo \citep{Suyu2014}. In this study, we construct dark and light mass models for 12 isolated strong lenses at the pixel level in order to explore their dark and luminous mass structure in detail.

The paper is organised as follows. In Section~\ref{sec:chap8sec2}, we introduce the data, the mass models and our lensing and dynamical modelling methods. Section~\ref{sec:chap8sec3} presents our main results, which we discuss in Section~\ref{sec:chap8sec4}. Finally, we summarise and conclude in Section~\ref{sec:chap8sec5}. Throughout the paper, we assume a flat $\Lambda$CDM cosmology with $\Omega_m = 0.3$ and $h = 0.7$.

\section{Data and modelling}
\label{sec:chap8sec2}

We construct simultaneous strong lensing and dynamical models for 12 of the EELs presented in \citet{Oldham2017a} (we exclude J1619, whose lensed features we were unable to fully reproduce using a parametric source model). To this end, we combine imaging and kinematic data as detailed below. 

\subsection{Data}
\label{sec:chap8sec2sub1}

The EELs were observed using HST/ACS (GO: 13661, PI: Auger) as presented in \citet{Oldham2017a}. Each EEL was observed at two dither positions for $\sim 500$ seconds per dither in the F555W/F606W (henceforth $V$-band) and F814W (henceforth $I$-band) filters -- with the former dependent on the lens redshift, and chosen to straddle the Balmer break. These observations were reduced using \textsc{Astrodrizzle} with a pixel scale of 0.05 arcsec pix$^{-1}$. We show postage stamps of these systems in Figure 1.

The EELs were also observed with ESI/Keck \citep{Sheinis2002}, and the spectra are presented in \citet{Oldham2017b}. Each EEL was observed for $\sim1$ hour using a 0.75 arcsec slit; the spectra were extracted using a custom-made \textsc{Python} code and the first and second velocity moments inferred by modelling the lens and source components simultaneously using stellar templates from the \textsc{INDO-US} library, as described in \citet{Oldham2017b}. For this study,  we extract kinematics for the lens galaxies over rectangular apertures extending 1 arcsec either side of the lens such that they probe the mass beyond the Einstein radius  (which is $\sim 0.5$ arcsec for a typical EEL; see Table 2 of \citealp{Oldham2017a}). 

\subsection{Mass model}
\label{sec:chap8sec2sub2}

We combine these imaging and kinematic datasets to construct a model in which we are able to disentangle the dark and luminous contributions to the total mass profile of each EEL. For the main analysis, we consider the mass profile that is detailed below; however, in Section~\ref{sec:chap8sec4}, we also explore alternative models in order to investigate the robustness and limitations of our inference.

We treat the mass density $\rho_{tot}(r)$ of the lens galaxy as the sum of a dark matter halo and a stellar component:
\begin{equation}
\rho_{tot}(r) = \rho_{DM}(r) + \rho_{\star}(r).
\end{equation}
We do not include a black hole because realistic black hole masses are orders of magnitude smaller than the total Einstein mass and therefore have undetectable effects on the lensed features. Indeed, we investigate the effects of neglecting the black hole in the Appendix and show that it has a negligible impact on the inferred halo and stellar mass parameters.

To allow for astrophysical changes to the halo structure relative to the dark-matter-only NFW prediction, we model the halo as a modified gNFW profile:
\begin{equation}
\rho_{DM}(r) = \rho_0\Big(\frac{r}{r_s}\Big)^{-\gamma} \Big(1 + \frac{r^2}{r_s^2}\Big)^{\frac{\gamma-3}{2}}
\end{equation}
which is characterised by a scale radius $r_s$, inner slope $\gamma$ and mass scale $\rho_0$. The quadrature in the final bracket relative to the standard form for the gNFW profile was introduced by \citet{Munoz2001} to make the calculation of lensing deflection angles analytic, and only modifies the halo shape near the scale radius (where the profile transitions more sharply); our data only probe the mass at much smaller radii. To minimise degeneracies between $\rho_0$, $\gamma$ and $r_s$, we reparameterise the halo so that the mass scale is specified by M$_{\textrm{DM}}($R$<2.5\textrm{kpc})$, the projected dark matter mass within a circular 2.5 kpc aperture. We additionally assume that the halo is spherical and concentric with the stellar mass and light (though we also investigate the limitations of these assumptions in the Appendix).

We parameterise the stellar mass with a spatially constant stellar-mass-to-light ratio $\Upsilon_{\star}$ such that the stellar surface mass density $\Sigma_{\star}(R)$ is related to the stellar surface brightness $I_{\star}(R)$ as
\begin{equation}
\Sigma_{\star}(R) = \Upsilon_{\star} I_{\star}(R)
\end{equation}
with $I_{\star}(R)$ described by either one or two S\'ersic profiles and fixed by the inference in \citet{Oldham2017a}. The photometry of the lens galaxies is summarised in Table~\ref{tab:phot}. The 3D mass density is then obtained by deprojecting $\Sigma_{\star}(R)$ assuming axisymmetry. 

\begin{table*}
\centering
\begin{tabular}{cccccc}\\\hline
lens & $z$ & $m_V$ (mag) & $m_I$ (mag) & $m_K$ (mag) & $R_{e}$ (kpc) \\\hline
J0837 & 0.4248 &  $ 19.52_{-0.01}^{+0.01} $ &  $ 18.53_{-0.02}^{+0.02} $ & $ 17.58_{-0.49}^{+0.29} $ & $ 7.90_{-0.09}^{+0.10} $ \\ 
J0901 & 0.3108 &  $ 19.84_{-0.01}^{+0.01} $ &  $ 18.89_{-0.01}^{+0.01} $ & $ 17.43_{-0.04}^{+0.03} $ & $ 3.27_{-0.04}^{+0.03} $ \\
J0913 & 0.3946 &  $ 20.55_{-0.02}^{+0.02} $ &  $ 18.97_{-0.02}^{+0.01} $ & $ 18.41_{-0.26}^{+0.33} $ & $ 5.44_{-0.11}^{+0.08} $ \\
J1125 & 0.4419 &  $ 19.57_{-0.01}^{+0.01} $ &  $ 18.42_{-0.01}^{+0.01} $ & $ 17.44_{-0.01}^{+0.01} $ & $ 6.88_{-0.06}^{+0.09} $ \\
J1144 & 0.3715 &  $ 19.98_{-0.03}^{+0.03} $ &  $ 18.95_{-0.03}^{+0.03} $ & $ 17.38_{-0.03}^{+0.02} $ & $ 3.23_{-0.12}^{+0.12} $ \\
J1218 & 0.3177 &  $ 19.13_{-0.02}^{+0.01} $ &  $ 18.24_{-0.02}^{+0.02} $ & $ 17.26_{-0.03}^{+0.03} $ & $ 7.82_{-0.18}^{+0.22} $ \\
J1323 & 0.3192 &  $ 19.95_{-0.01}^{+0.01} $ &  $ 18.50_{-0.01}^{+0.01} $ & $ 17.35_{-0.52}^{+0.33} $ & $ 6.81_{-0.02}^{+0.03} $ \\
J1347 & 0.3974 &  $ 20.03_{-0.01}^{+0.02} $ &  $ 18.95_{-0.01}^{+0.02} $ & $ 17.84_{-0.05}^{+0.02} $ & $ 4.80_{-0.11}^{+0.13} $ \\
J1446 & 0.3175 &  $ 19.83_{-0.01}^{+0.01} $ &  $ 18.93_{-0.01}^{+0.01} $ & $ 17.51_{-0.06}^{+0.07} $ & $ 3.89_{-0.02}^{+0.04} $ \\
J1605 & 0.3065 &  $ 19.70_{-0.01}^{+0.01} $ &  $ 18.29_{-0.01}^{+0.01} $ & $ 17.14_{-0.14}^{+0.15} $ & $ 5.08_{-0.03}^{+0.03} $ \\
J1606 & 0.3810 &  $ 19.36_{-0.01}^{+0.01} $ &  $ 18.32_{-0.01}^{+0.01} $ & $ 17.73_{-0.01}^{+0.01} $ & $ 5.99_{-0.10}^{+0.10} $ \\
J2228 & 0.2387 &  $ 19.13_{-0.01}^{+0.01} $ &  $ 17.81_{-0.01}^{+0.01} $ & $ 17.22_{-0.05}^{+0.07} $ & $ 7.74_{-0.09}^{+0.09} $ \\\hline
\end{tabular}
\caption{Photometry for the 12 EELs lens galaxies used in this study. Columns specify: lens name, lens redshift $z$, apparent AB magnitudes in the $V$, $I$ and $K$ bands, and total effective radius $R_{e}$. These measurements are based on the inference presented in \citet{Oldham2017a}, in which the lens and source galaxy light distributions were modelled as either one or two S\'ersic components, and the lens mass distribution was modelled as a power law. When two S\'ersic components were required, the effective radius given is defined as the radius within which half the total light (summed from the two components) is contained. The uncertainties given are statistical uncertainties from the inferences (i.e. not accounting for any systematics). Redshifts are from \citet{Oldham2017b}, based on Keck/ESI spectra.}
\label{tab:phot}
\end{table*}

\subsection{Lens modelling}
\label{sec:chap8sec2sub3}

We make full lensing reconstructions of the HST/ACS imaging using an extension of the methods presented in \citet{Oldham2017a}. Using the results of that study, which described the total mass with a power-law profile and inferred the light distributions of the source and lens simultaneously, we subtract the flux contribution of the lens; we then parameterise the source light using either one or two S\'ersic profiles, and calculate the deflection angles based on the mass models presented in Section~\ref{sec:chap8sec2sub2} to form the source in the image plane. We also allow for an external shear to account for perturbations due to additional masses along the line of sight, described by a magnitude $\Gamma$ and direction $\theta_{\Gamma}$. We use an unsaturated star in the HST field as a point-spread function (PSF), which we convolve with the model image to allow comparison with the data. We note that, in \citet{Oldham2017a}, we tested a subset of systems to verify that the choice of PSF star did not have any significant influence on our inference. Since the HST imaging used here is identical to that study, we do not repeat the test here. For a given set of non-linear parameters describing the lens mass and source light profiles, we then determine the best combination of amplitudes for the source light and a uniform background component using a least-squares linear inversion. We model the $V$ and $I$ bands simultaneously (i.e. dictating that the luminous source structure is the same in both bands), allowing for a spatial offset between them due to image registration uncertainties, and calculate the contribution to the likelihood of the data $\vec{D}$ given the model $\vec{M}$ as
\begin{equation}
\ln L(\vec{D}|\vec{M}) = -\frac{1}{2} \sum_i \Big(\frac{d_i-m_i}{n_i}\Big)^2
\end{equation}
where $d_i$, $m_i$ and $n_i$ are the $i^{th}$ pixel in the data image, model image and noise map respectively, and the sum is over all unmasked pixels (for some systems, bright interloping objects are masked by hand). We investigate the possibility of also inferring a weighting term $\beta_f$ for each filter such that $n_{i,f} = \beta_f n_i$, to allow the model to maximise the combined posterior of the two lensing datasets and the dynamical data; however, we find that this generally leads to an overfitting of the kinematics, and we do not include these parameters in our final modelling. In Appendix A, we also construct pixellated models in order to verify that our inference on the lens mass structure does not depend on assumptions about the luminous structure of the source.

\subsection{Dynamical modelling}
\label{sec:chap8sec2sub4}

For a given mass model, we use the spherical Jeans equation \citep[see e.g.][]{Mamon2005} to calculate the stellar velocity dispersion within a circular aperture of radius 1 arcsec assuming isotropic orbits. We investigate the possibility of allowing the anisotropy to take a non-zero but spatially constant value, but find that our data are not sufficient to constrain this extra parameter (see Appendix A). We also note that the spherical assumption is an approximation given that the EELs lenses have light distributions that are described by elliptical S\'ersic profiles; however, \citet{Sonnenfeld2015} showed that approximating a S\'ersic profile with $q = 0.85$ using the spherical Jeans equation in this way only gives rise to an uncertainty of a few kms$^{-1}$.

The likelihood of the observed velocity dispersion given the model is then measured in a chi-squared sense as 
\begin{equation}
\ln(\vec{D}|\vec{M}) = -\frac{1}{2} \Big(\frac{\sigma_d - \sigma_m}{\sigma_n}\Big)^2
\end{equation}
for observed velocity dispersion $\sigma_d$, model velocity dispersion $\sigma_m$ and uncertainty $\sigma_n$. We note that, although there are many lens image pixels and only one velocity dispersion measurement, the latter still plays an important role in the inference by guiding it away from models with unreasonably large or small central 3D masses.

We combine dynamical and lensing likelihood terms to explore the posterior probability distribution of the model given the data using Markov Chain Monte Carlo (MCMC) sampling, as implemented in \textsc{Emcee} \citep{ForemanMackey2013}.

\begin{figure*}
\centering
\includegraphics[trim=110 20 110 20,clip,width=0.24\textwidth]{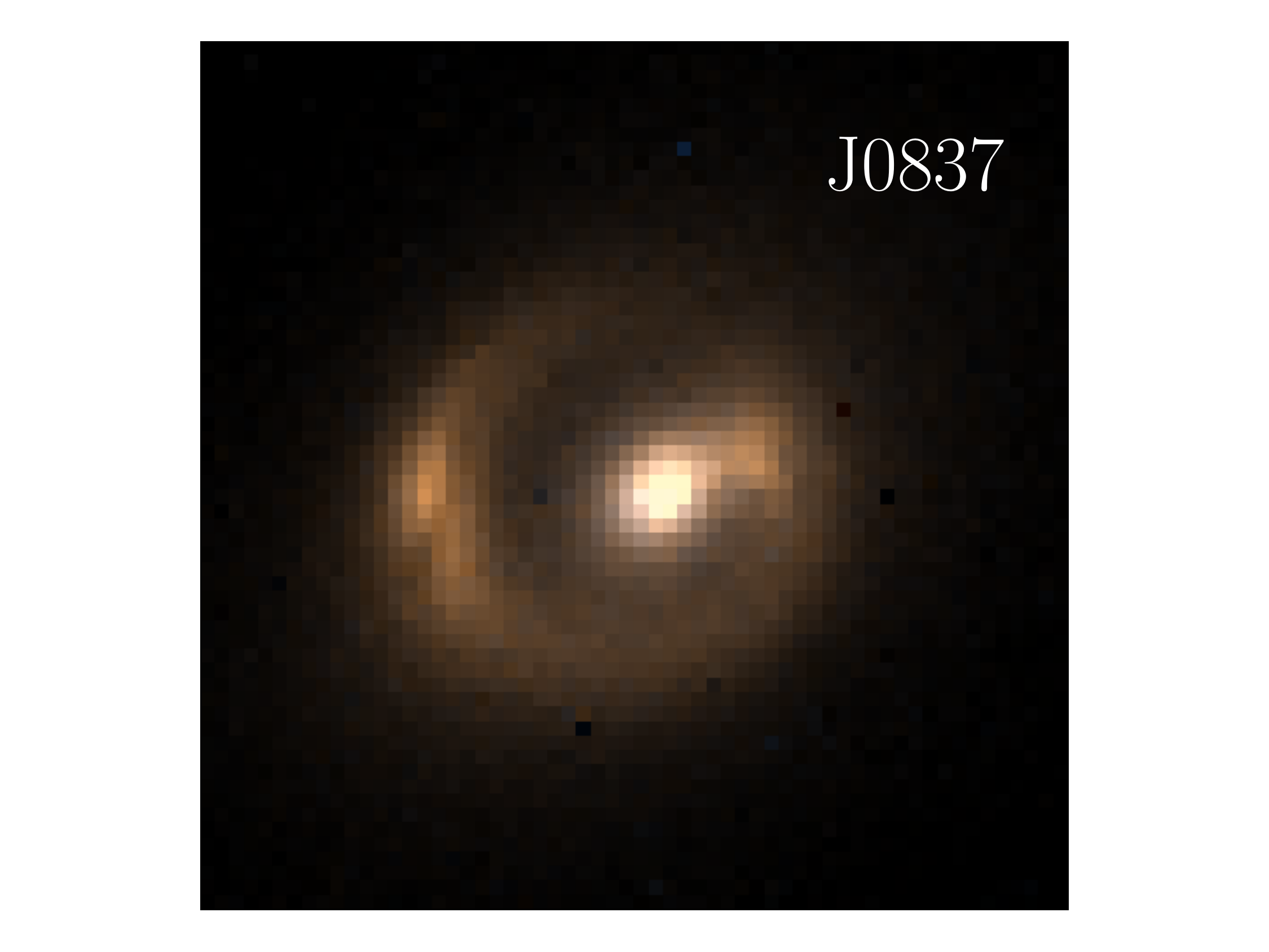}
\includegraphics[trim=110 20 110 20,clip,width=0.24\textwidth]{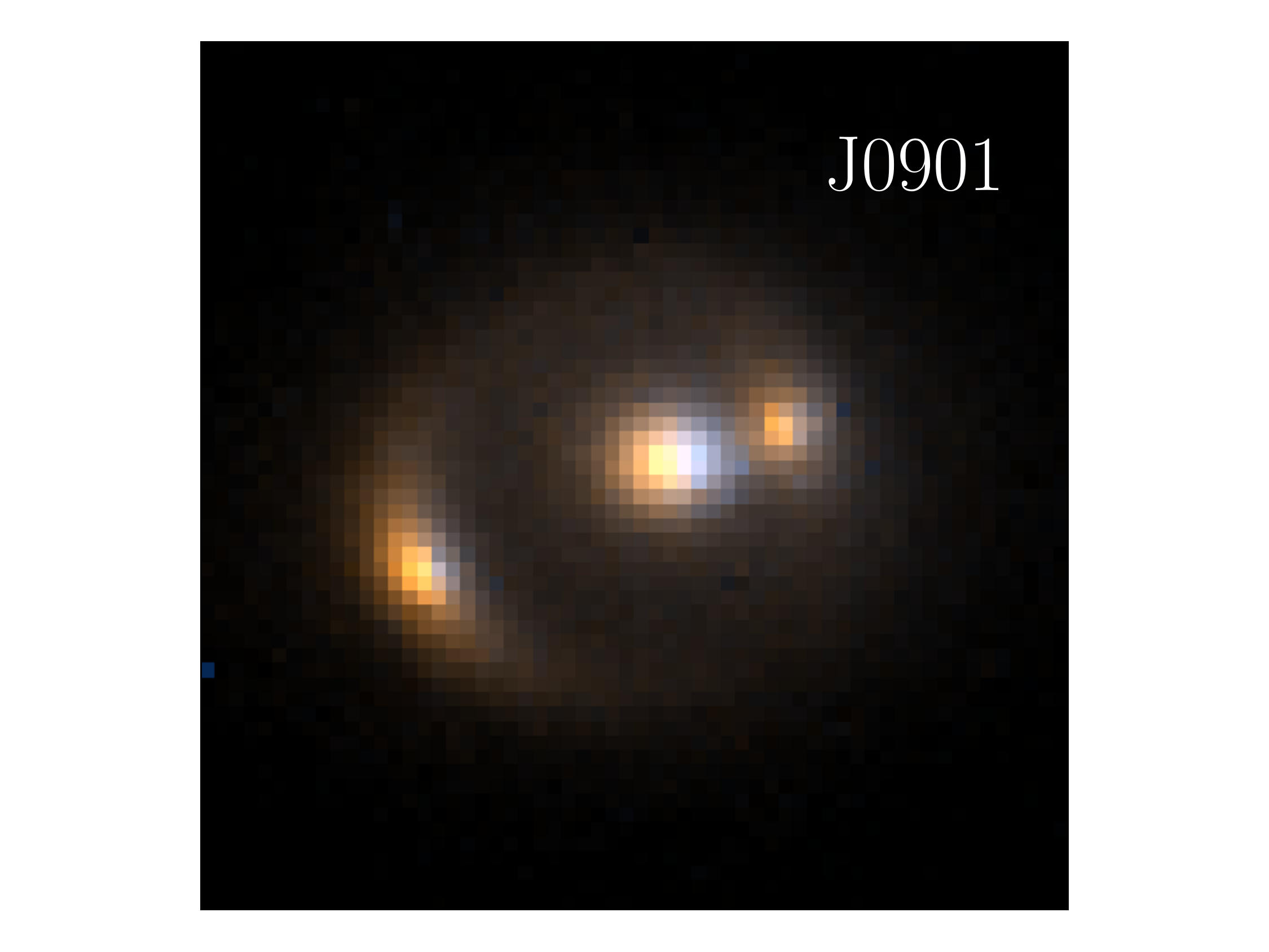}
\includegraphics[trim=110 20 110 20,clip,width=0.24\textwidth]{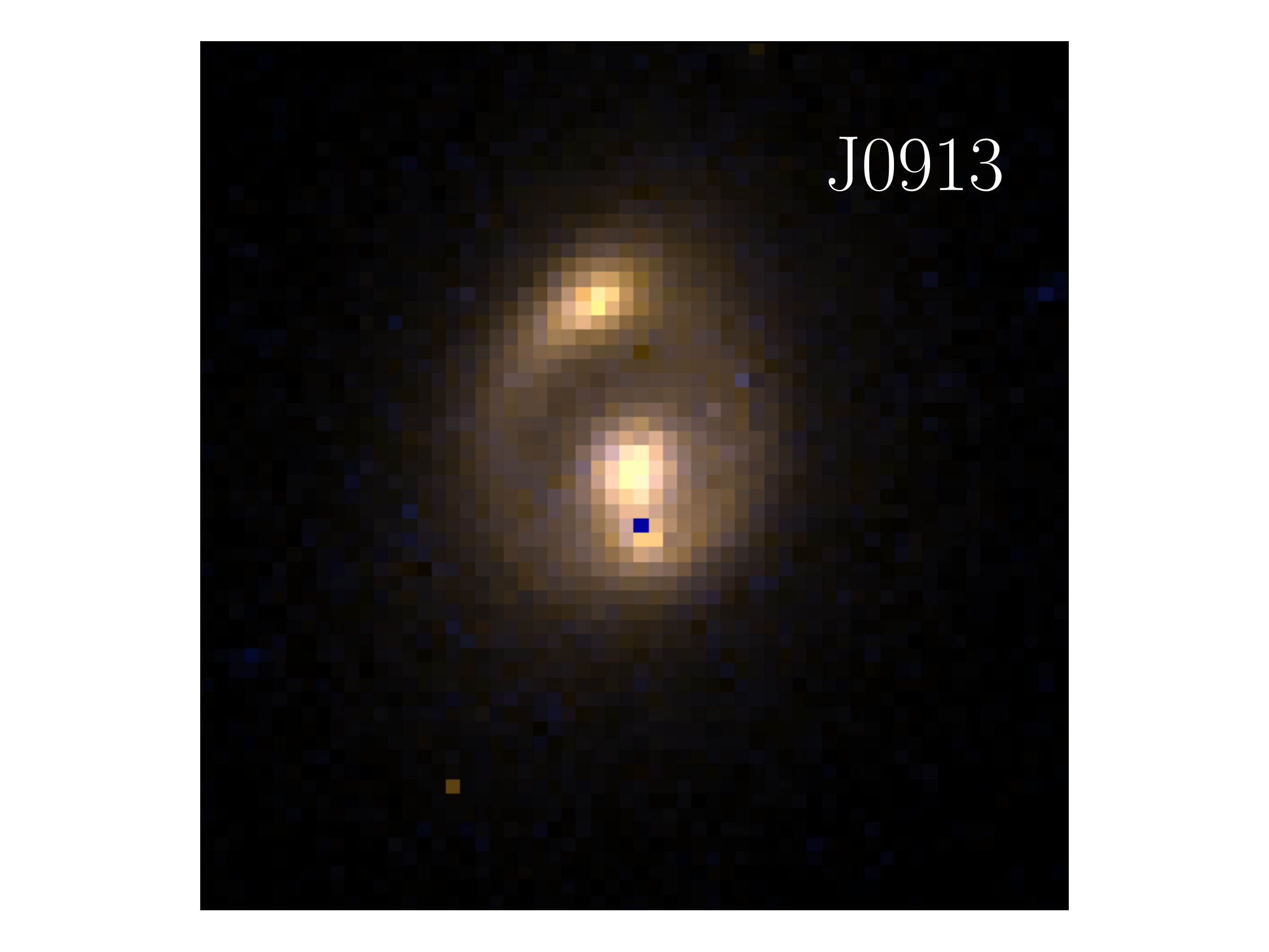}
\includegraphics[trim=110 20 110 20,clip,width=0.24\textwidth]{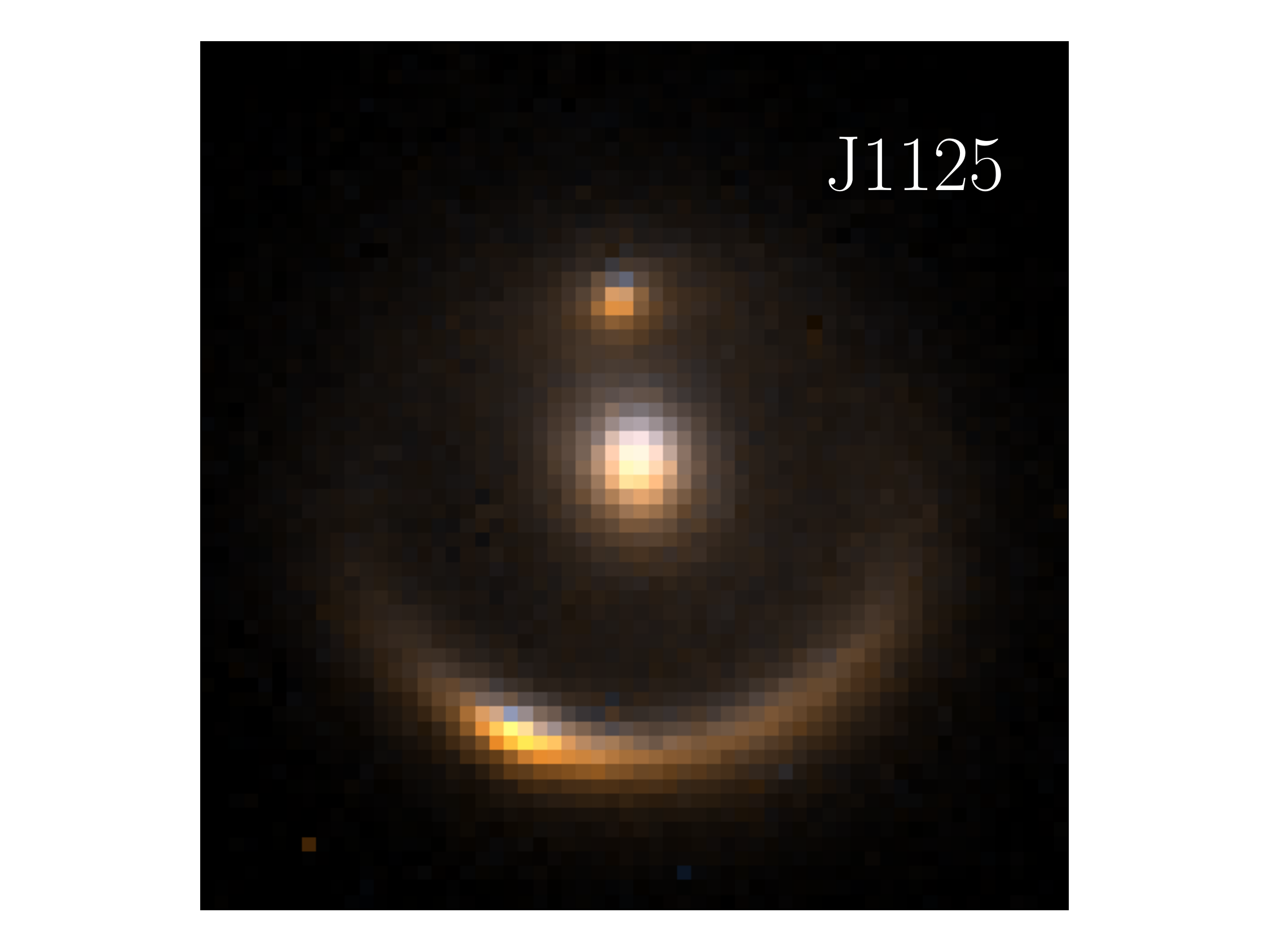}
\includegraphics[trim=110 20 110 20,clip,width=0.24\textwidth]{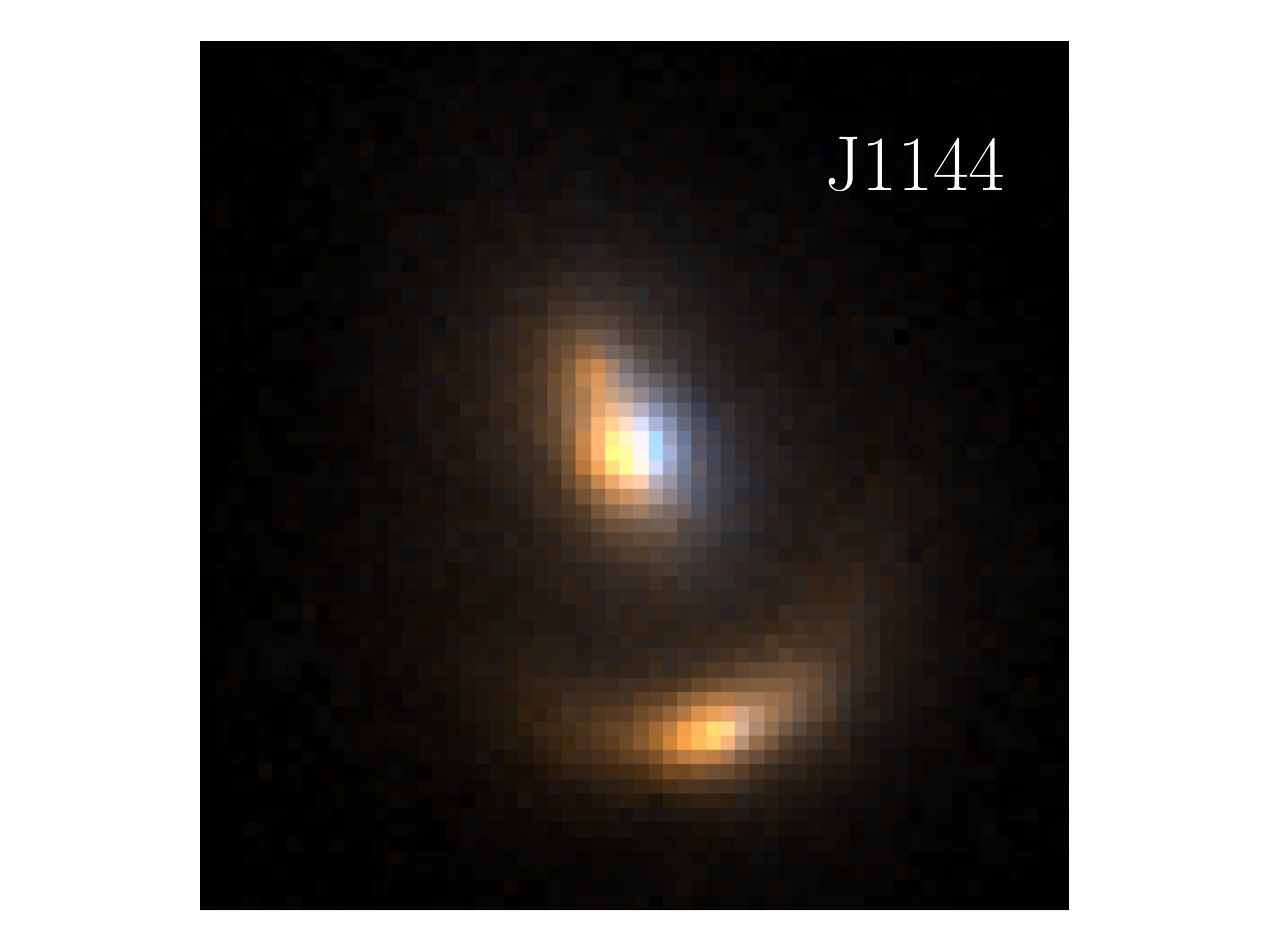}
\includegraphics[trim=110 20 110 20,clip,width=0.24\textwidth]{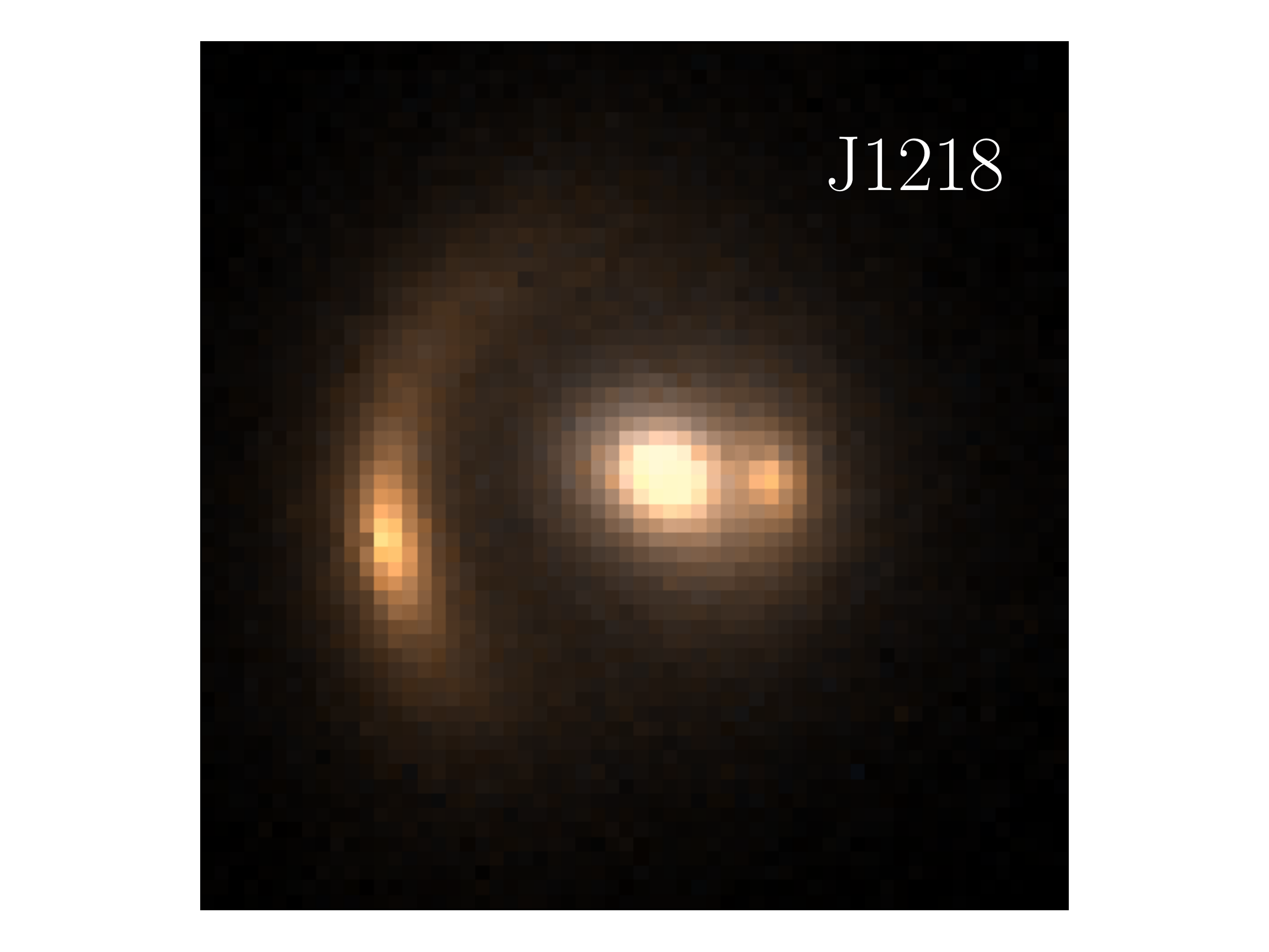}
\includegraphics[trim=110 20 110 20,clip,width=0.24\textwidth]{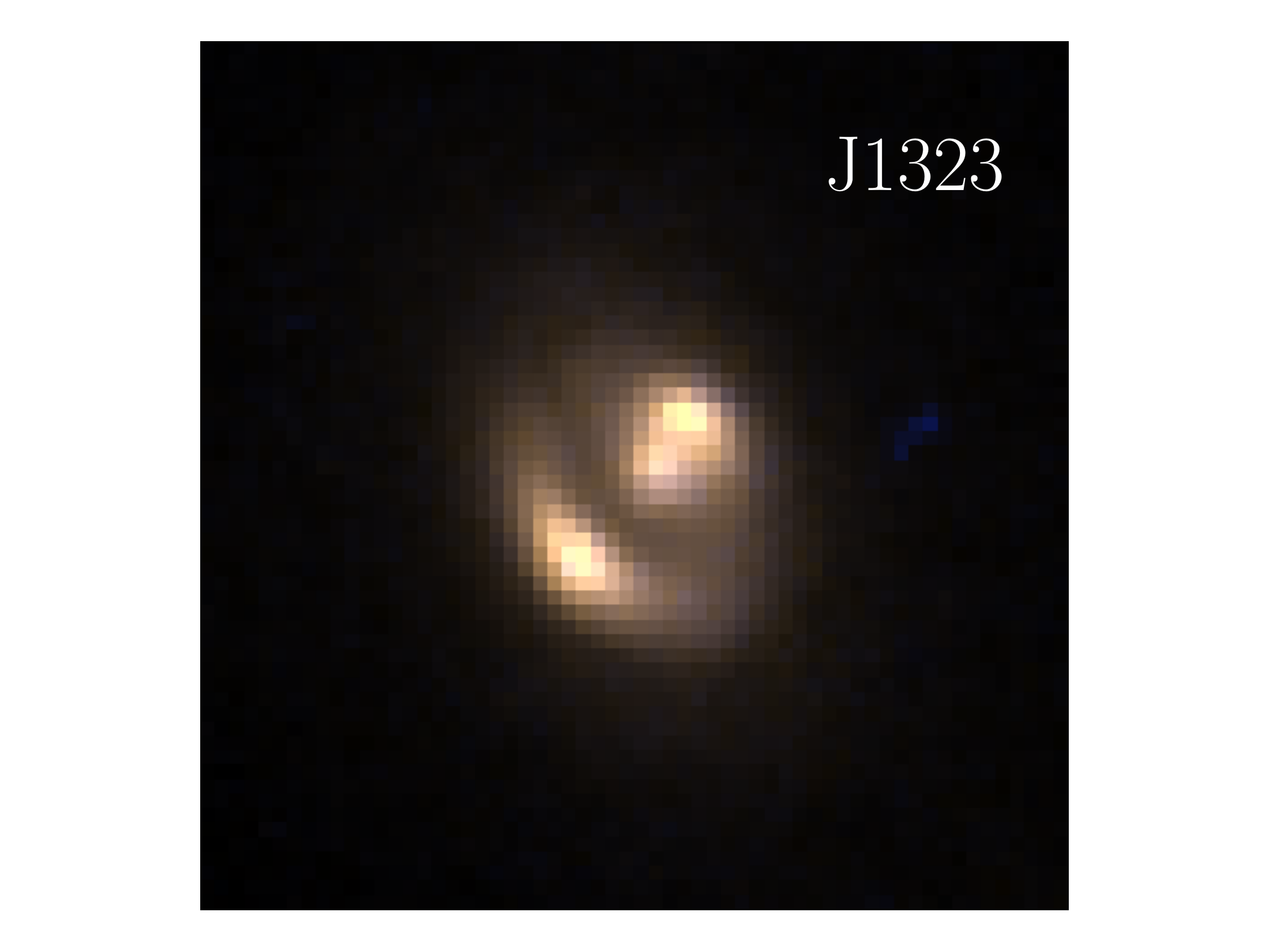}
\includegraphics[trim=110 20 110 20,clip,width=0.24\textwidth]{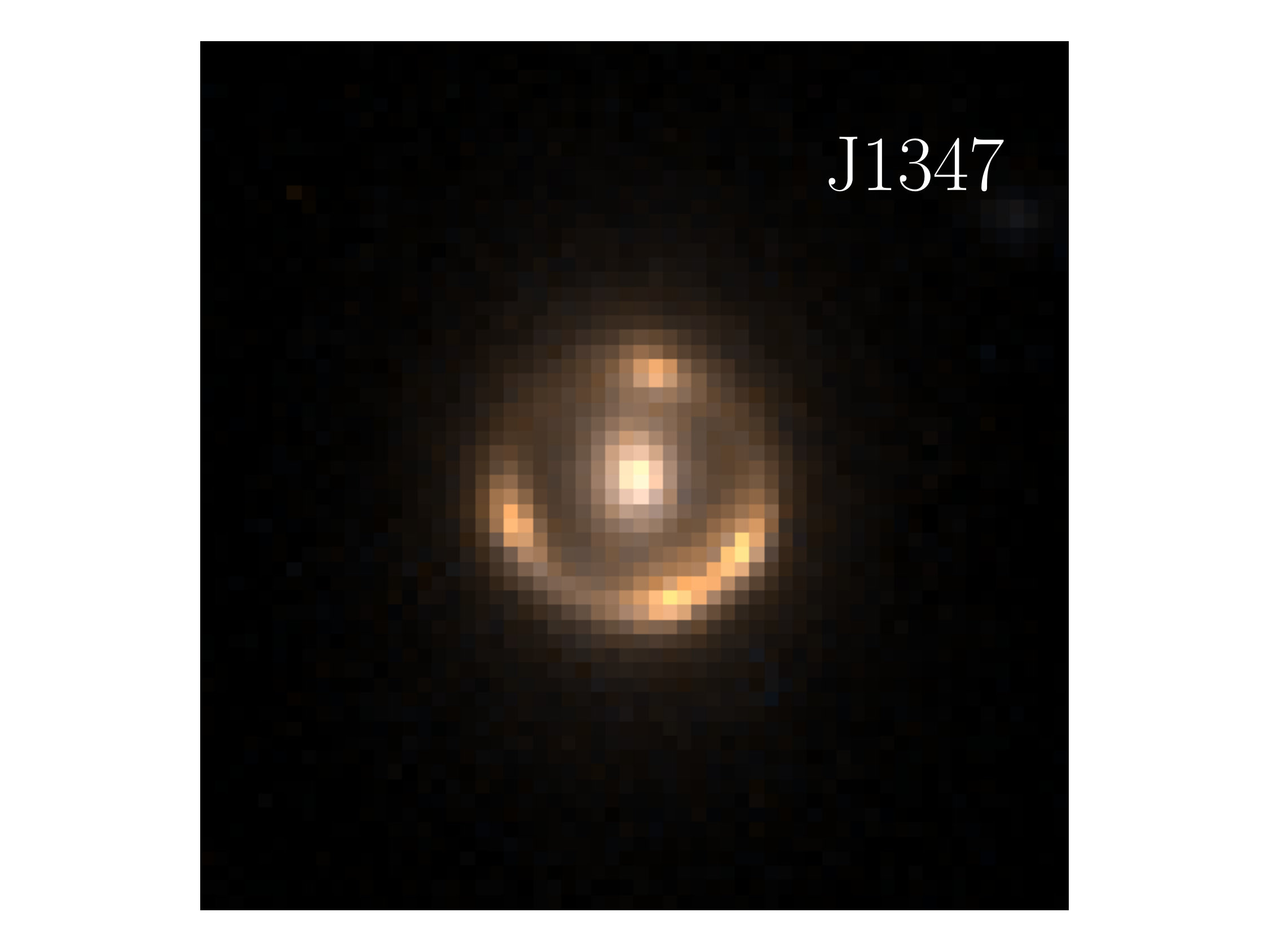}
\includegraphics[trim=110 20 110 20,clip,width=0.24\textwidth]{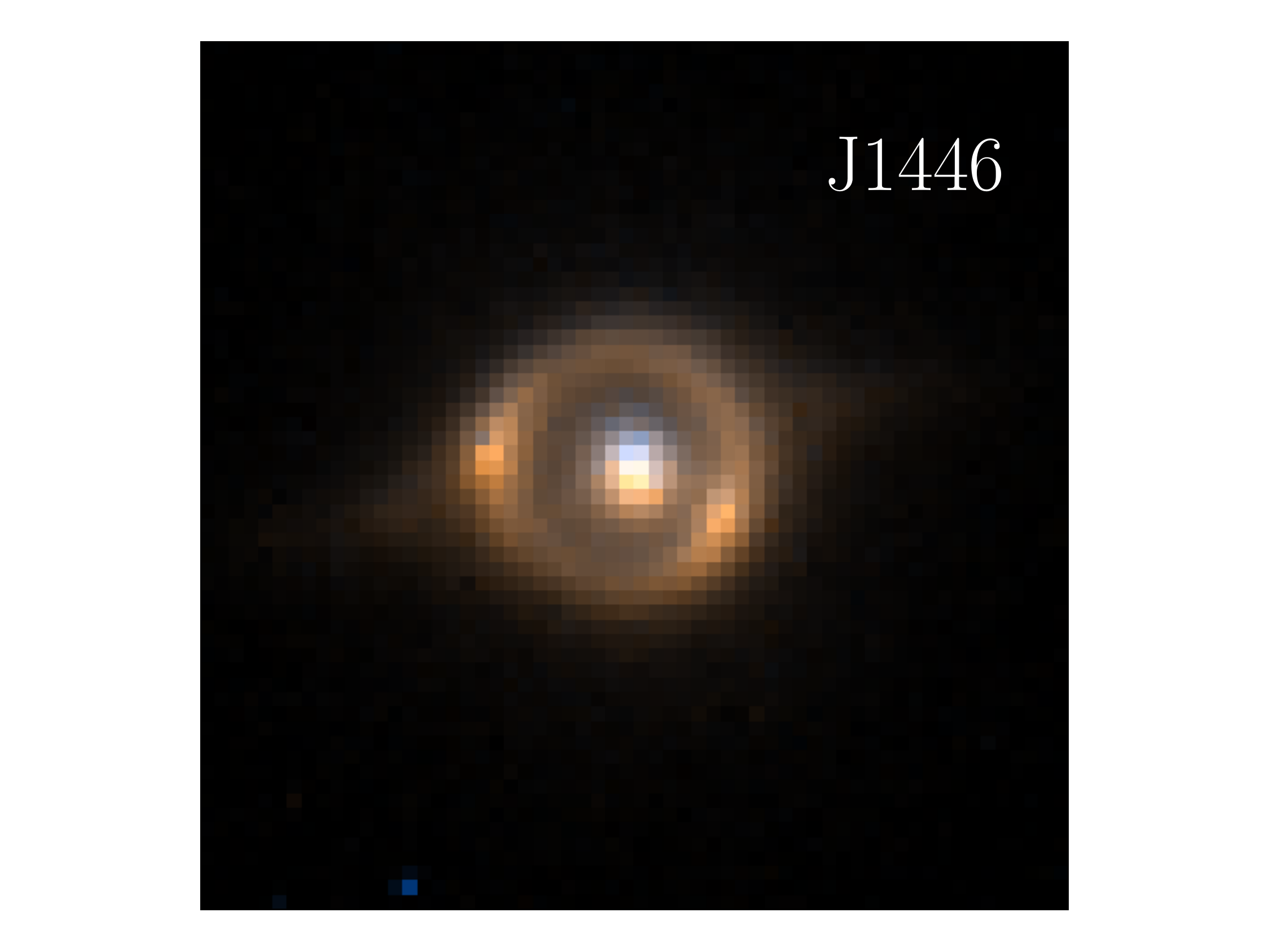}
\includegraphics[trim=110 20 110 20,clip,width=0.24\textwidth]{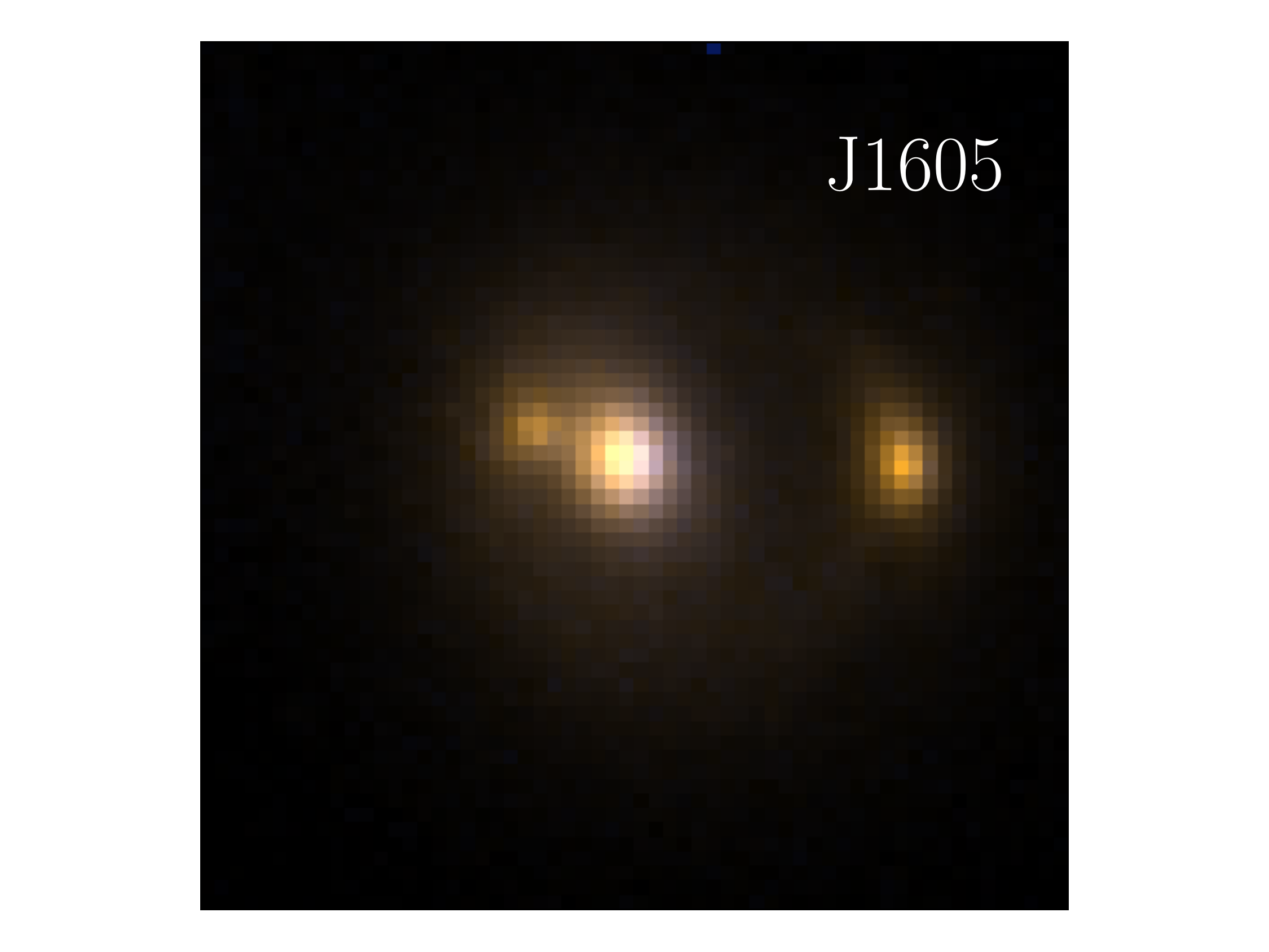}
\includegraphics[trim=110 20 110 20,clip,width=0.24\textwidth]{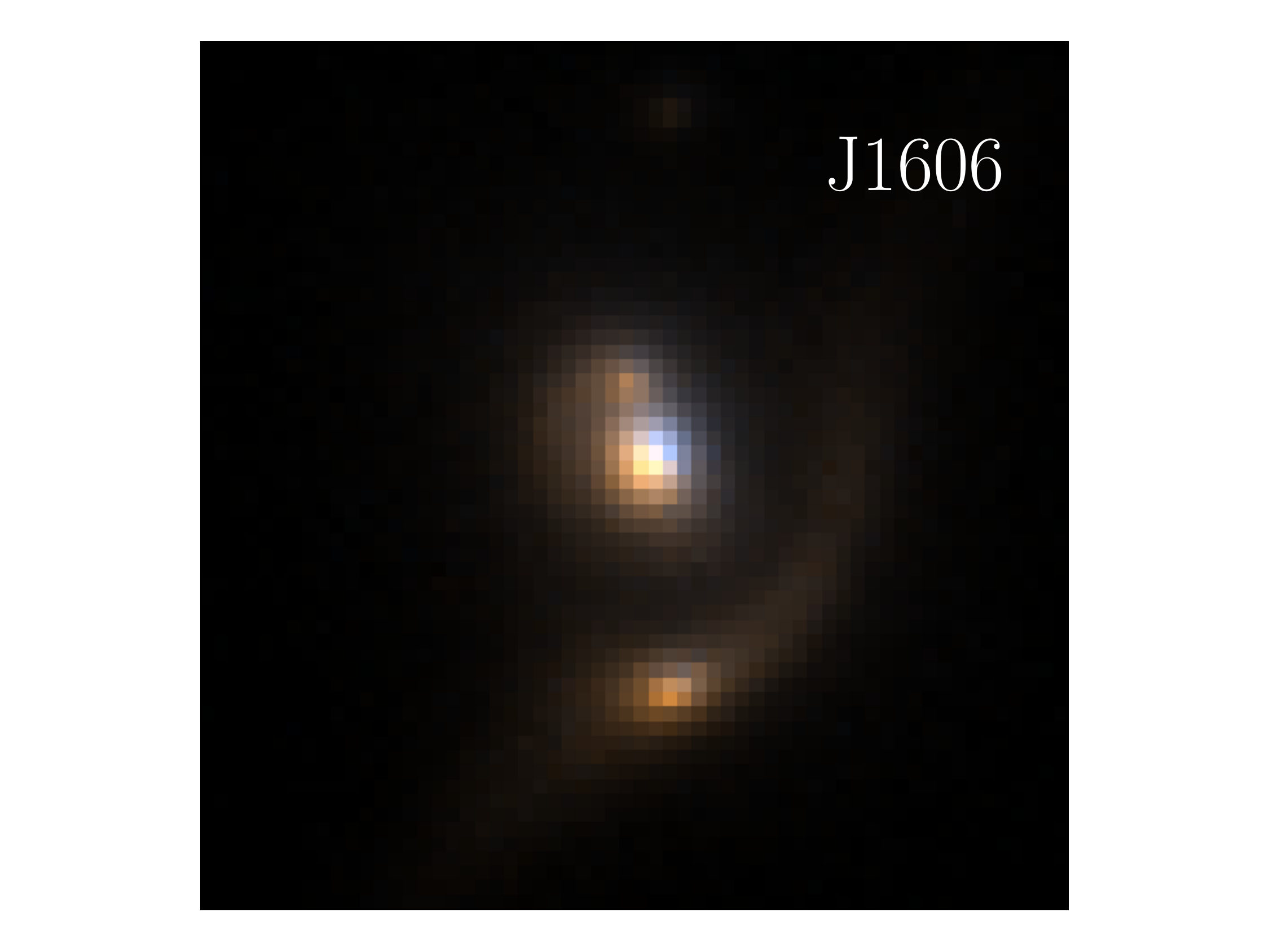}
\includegraphics[trim=110 20 110 20,clip,width=0.24\textwidth]{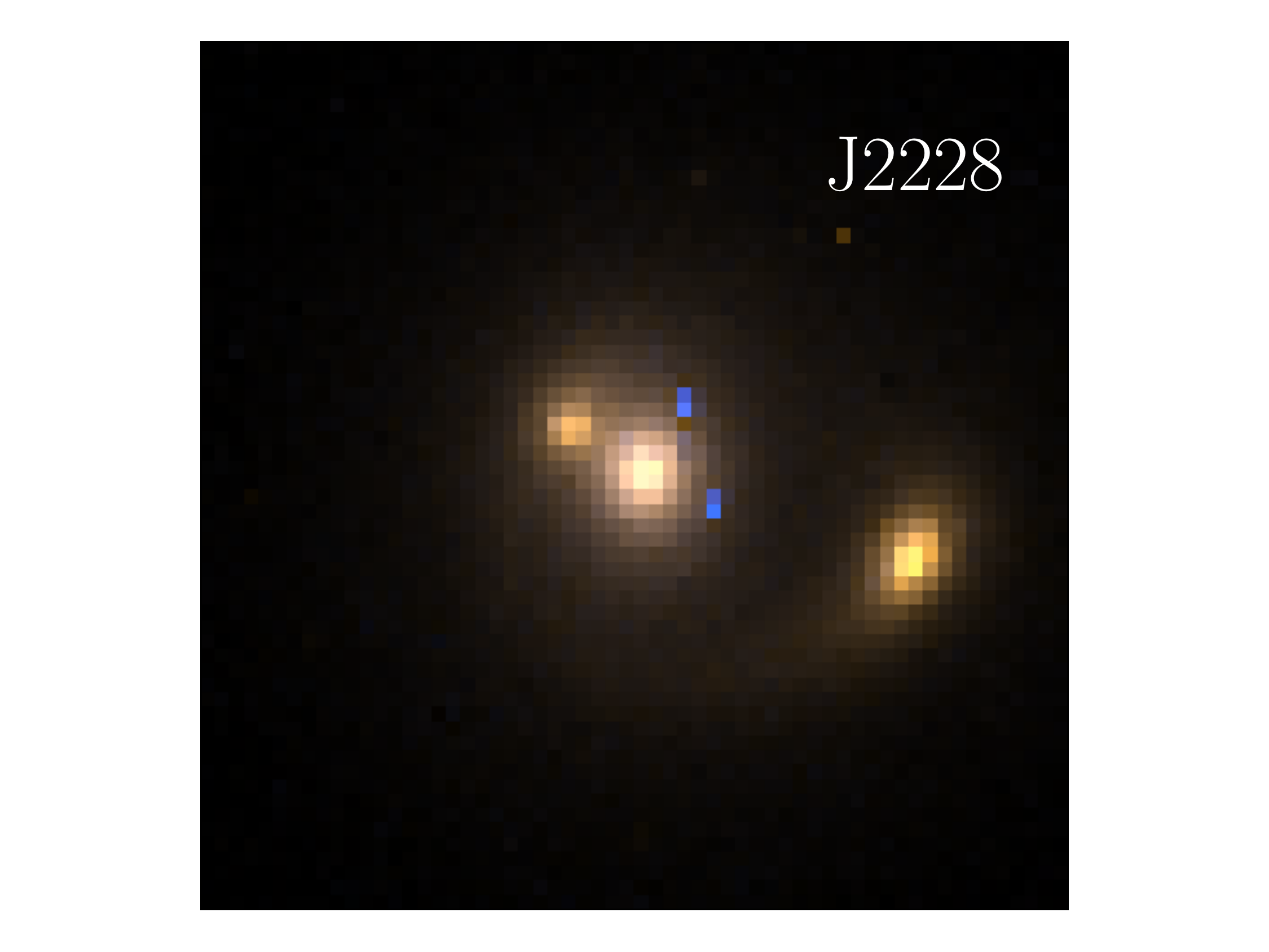}
\caption{HST VI images of the 12 EELs studied in this paper. The name used to refer to each is given in the top right-hand corner of each image.} 
\label{fig:postagestamps}
\end{figure*}

\begin{figure*}
\centering
\includegraphics[trim=30 60 60 60,clip,width=0.54\textwidth]{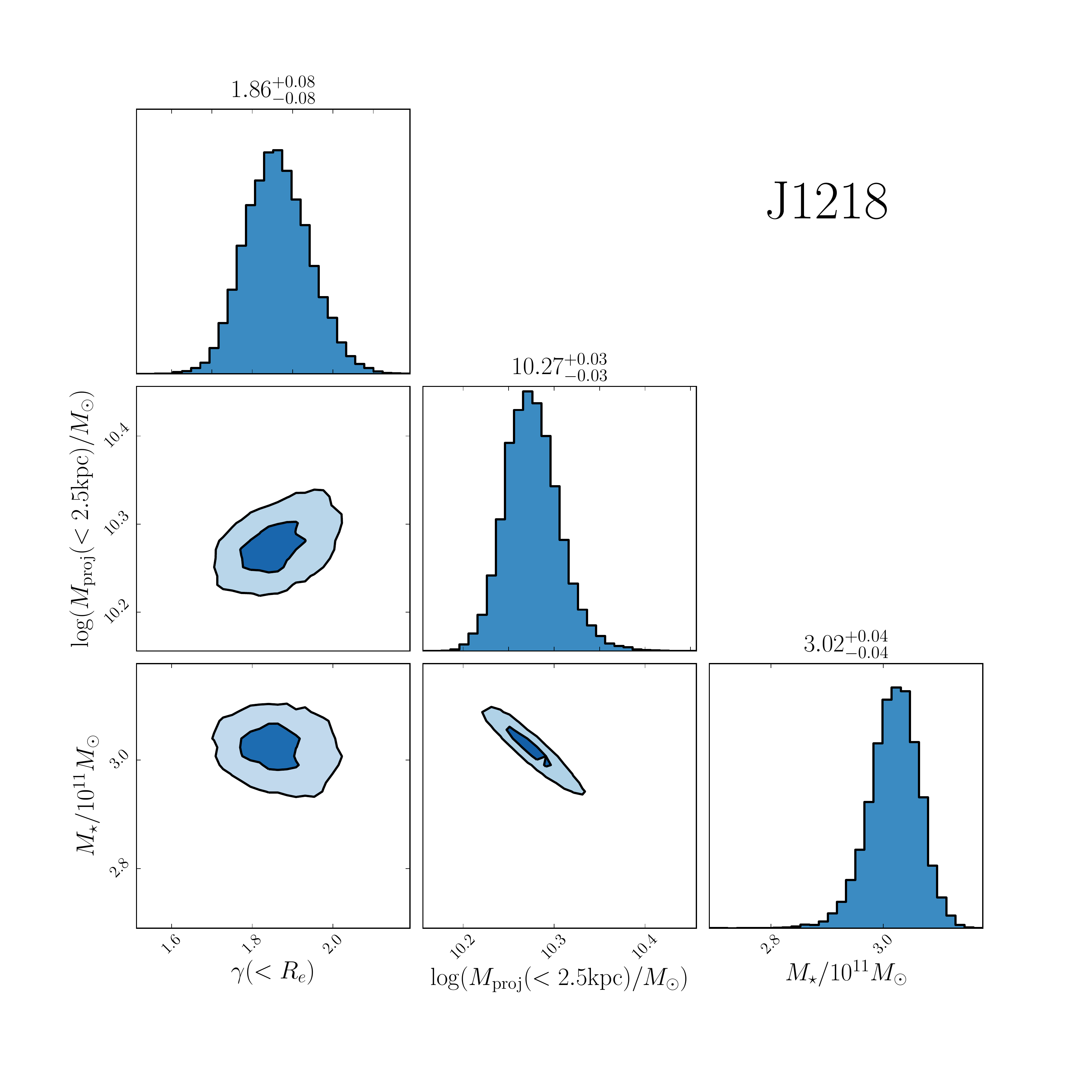}\hfill
%
\raisebox{1.2cm}{\includegraphics[trim=10 10 10 10,clip,width=0.44\textwidth]{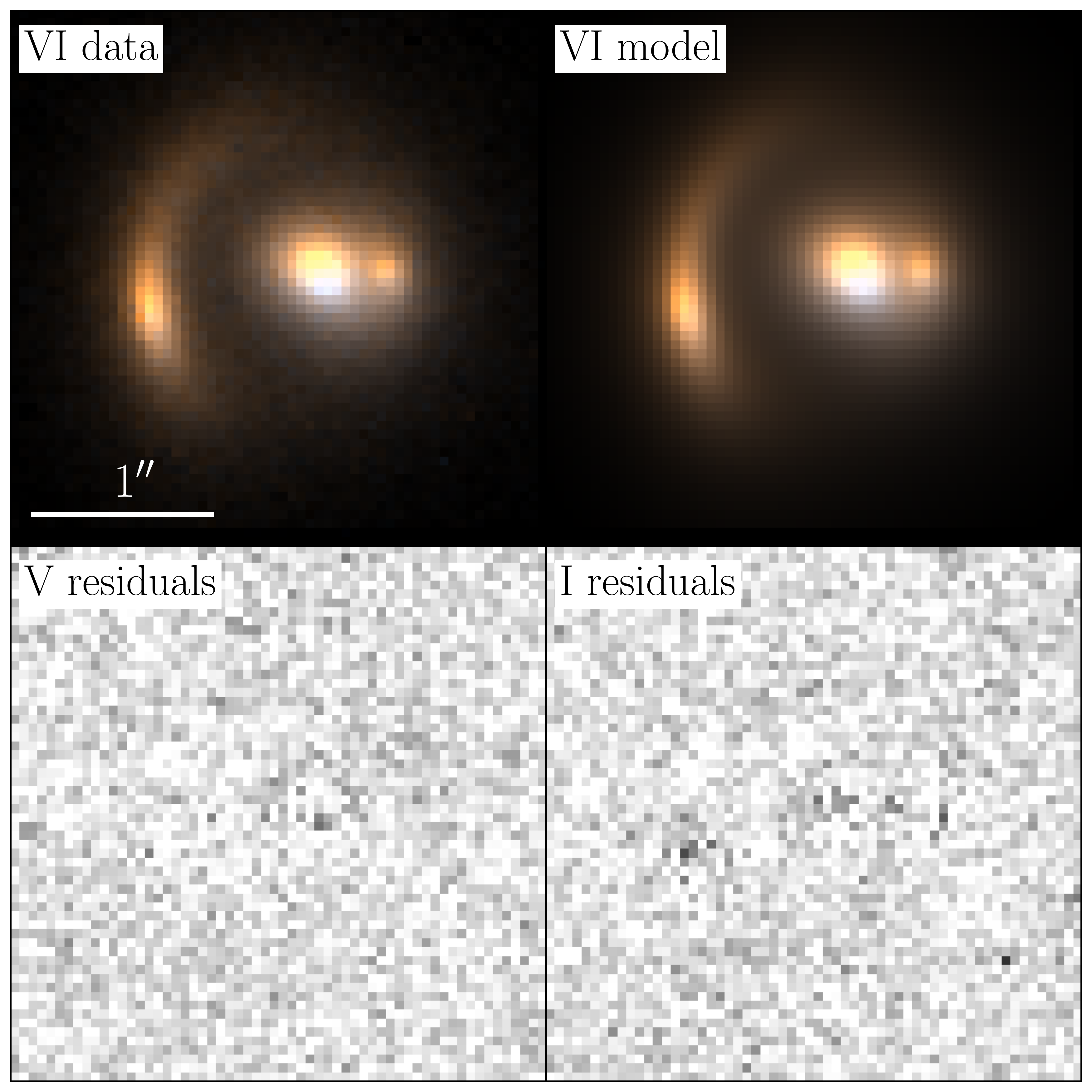}}\hfill
\caption{Left: Summary of our inference on the mass model parameters for a typical EEL (J1218). We show our 2D and 1D marginalised inferences on the mean mass-weighted halo slope within the effective radius $\gamma(<R_e)$, the projected dark matter mass within 2.5 kpc, $\log (M_{DM}/M_{\odot})$, and the total stellar mass $M_{\star}/10^{11}M_{\odot}$, and give the median values and 68\% confidence intervals of each above the 1D histograms. Right: Reconstructed image for the same EEL. We use the $V$ and $I$ HST images (top left) to make inference on the dark and luminous mass structure of the lens; our best model (top right) describes the data virtually down to the noise (bottom). }
\label{fig:chap8fig2}
\end{figure*}

\section{Results}
\label{sec:chap8sec3}

Figure 2 shows our inference on the mass parameters for a typical lens (J1218; left-hand panel), and the image, model and residuals for the same system (right). Table 2 presents our inferences on the mass structure for the full lens sample. In the following subsections, we present these results in detail.

\subsection{Stellar mass}
\label{sec:chap8sec3sub1}

The stellar-mass-to-light ratio $\Upsilon_{\star}$ inferred from lensing and dynamics gives a stellar mass $M_{\star,LD} = \Upsilon_{\star}I_{\star}(R)$, which we use to test for deviations of the stellar populations from a Milky-Way-like IMF. We use the stellar masses $M_{\star,chab}$ presented in \citet{Oldham2017a}, which were calculated via stellar population modelling of the photometry under the assumption of a Chabrier IMF (but with metallicity, age, reddening and the time constant of an exponentially decaying star formation history as free parameters), and calculate the IMF mismatch parameter
\begin{equation}
\alpha_c = \frac{M_{\star,LD}}{M_{\star,chab}}.
\end{equation}
Note that, with this definition, stellar populations with Chabrier and Salpeter IMFs have $\alpha_c = 1$ and $\alpha_c \sim 1.7$ respectively. The mismatch parameters are included in Table 2 and Figure 3. Note also that Figure 3 also shows $\alpha_{salp}$, which is normalised by the stellar mass inferred from the photometry assuming a Salpeter IMF; this is done to facilitate a comparison with the previous study of \citet{Sonnenfeld2015}.

We characterise the sample by modelling our inferences on $\alpha_c$ for individual EELs as being drawn from a Gaussian distribution $\alpha_c \sim N(\mu_{\alpha c},\tau_{\alpha c}^2)$. This constitutes a hierarchical model with hyperparameters $\omega = (\mu_{\alpha c},\tau_{\alpha c})$ whose posterior probability distribution, given the data $D$, is given by 
\begin{equation}
\begin{split}
&P(\omega|D) \propto P(\omega) P(D|\omega) \\
 = P(\omega) \prod_i & \int d\alpha_{c} P_i(D_i|\alpha_{c}) N(\alpha_{c}; \omega) \\
 \propto P(w) \prod_i & \int d\alpha_{c} \frac{P_i(\alpha_{c}|D_i)}{P_i(\alpha_{c})} N(\alpha_{c};\omega).
\end{split}
\label{eq7}
\end{equation}
The first term $P(\omega)$ is the prior probability on $\omega$, which we assume to be uniform; $P_i(\alpha_{c})$ is the prior on $\alpha_c$ for the $i^{th}$ EEL, which is uniform and the same for all EELs; $P_i(\alpha_{c}|D_i)$ is the posterior on $\alpha_c$ for the $i^{th}$ EEL as inferred from our lensing and dynamical modelling. Equation~\ref{eq7} is therefore a product of $i$ integrals of the trial parent distribution multiplied by the posterior on $\alpha_c$ for the $i^{th}$ lens, which we perform via Monte Carlo integration and sample with the same MCMC methods as in the previous section to find $\mu_{\alpha c} = 1.80 \pm 0.14$ (Table~\ref{tab:chap8tab3}); moreover, $\mu_{\alpha c} > 1.3$ with $99\%$ confidence. Under our model assumptions, then, the stellar masses of the EELs lens sample strongly favour Salpeter-like IMFs over Chabrier ones, consistent with previous results from lensing and dynamics \citep[e.g.][]{Auger2010b, Cappellari2012}. We compare our inference with expectations assuming Chabrier and Salpeter IMFs, and with the SLACS results from \citet{Auger2010b}, in Figure 3 (left).


We also note that our sample exhibits a large scatter in $\alpha_c$ -- with $\tau_{\alpha c} = 0.49_{-0.10}^{+0.14}$ -- in spite of its small range in $\sigma_{\star}$ (Figure 3, right). Whilst some previous lensing studies assuming more rigid forms for the dark matter structure (i.e. NFW haloes; \citealp{Treu2010}) have found a positive correlation between $\alpha_c$ and $\sigma_{\star}$ for other ETG samples, \citet{Auger2010b} has shown that this trend is removed when haloes are allowed to be adiabatically contracted. The small range of $\sigma_{\star}$ across our sample prohibits an investigation into such trends in this study, but it is possible that the large $\alpha_c$ scatter in our sample is also due to the increased flexibility of our halo models. If true, this highlights the sensitivity of our constraints on the IMF to our ability to disentangle the dark and luminous mass components in these systems.

\begin{figure*}
\centering
\includegraphics[trim=10 10 10 10,clip,width=0.49\textwidth]{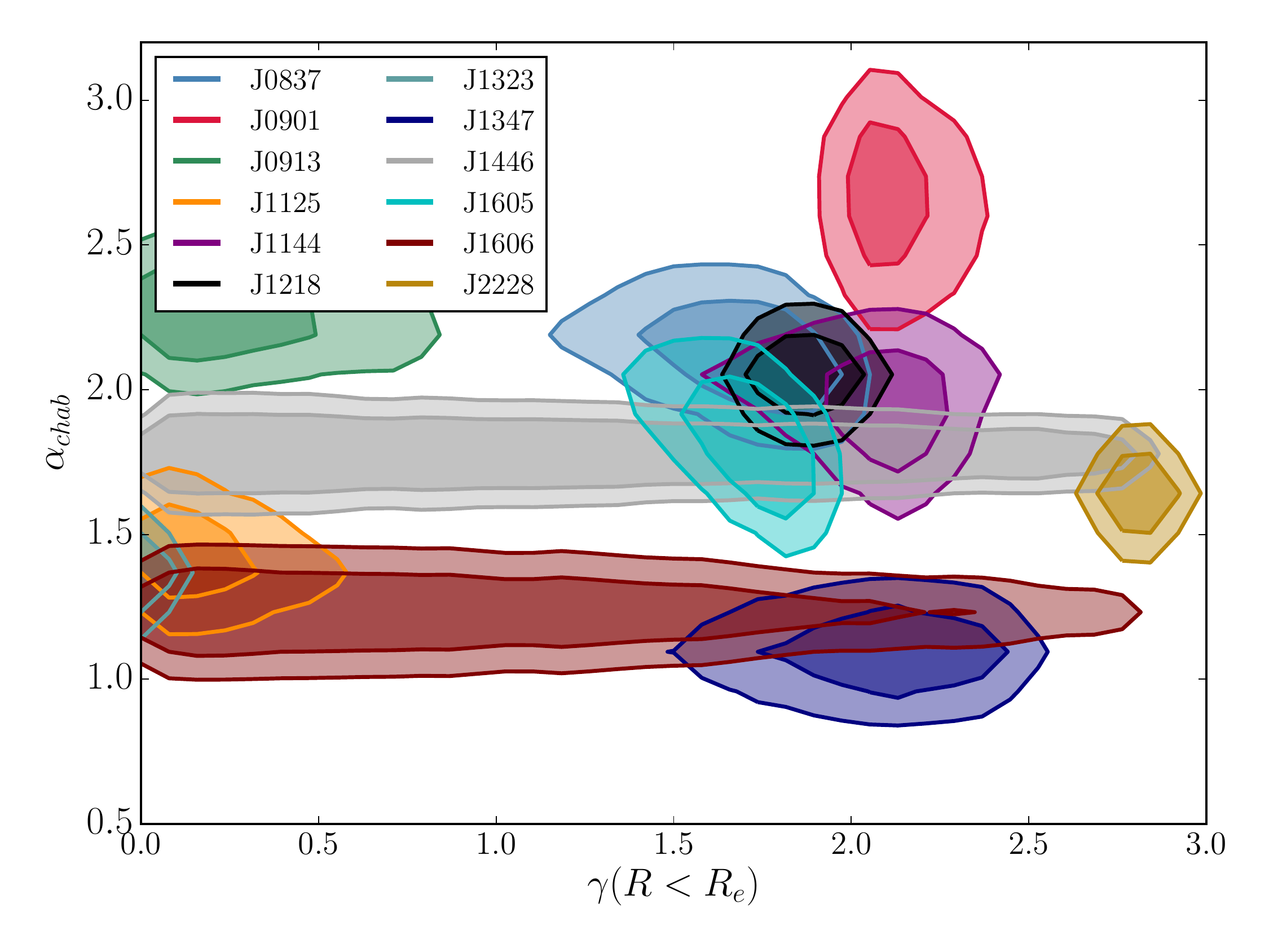}\hfill
\includegraphics[trim=10 10 10 10,clip,width=0.49\textwidth]{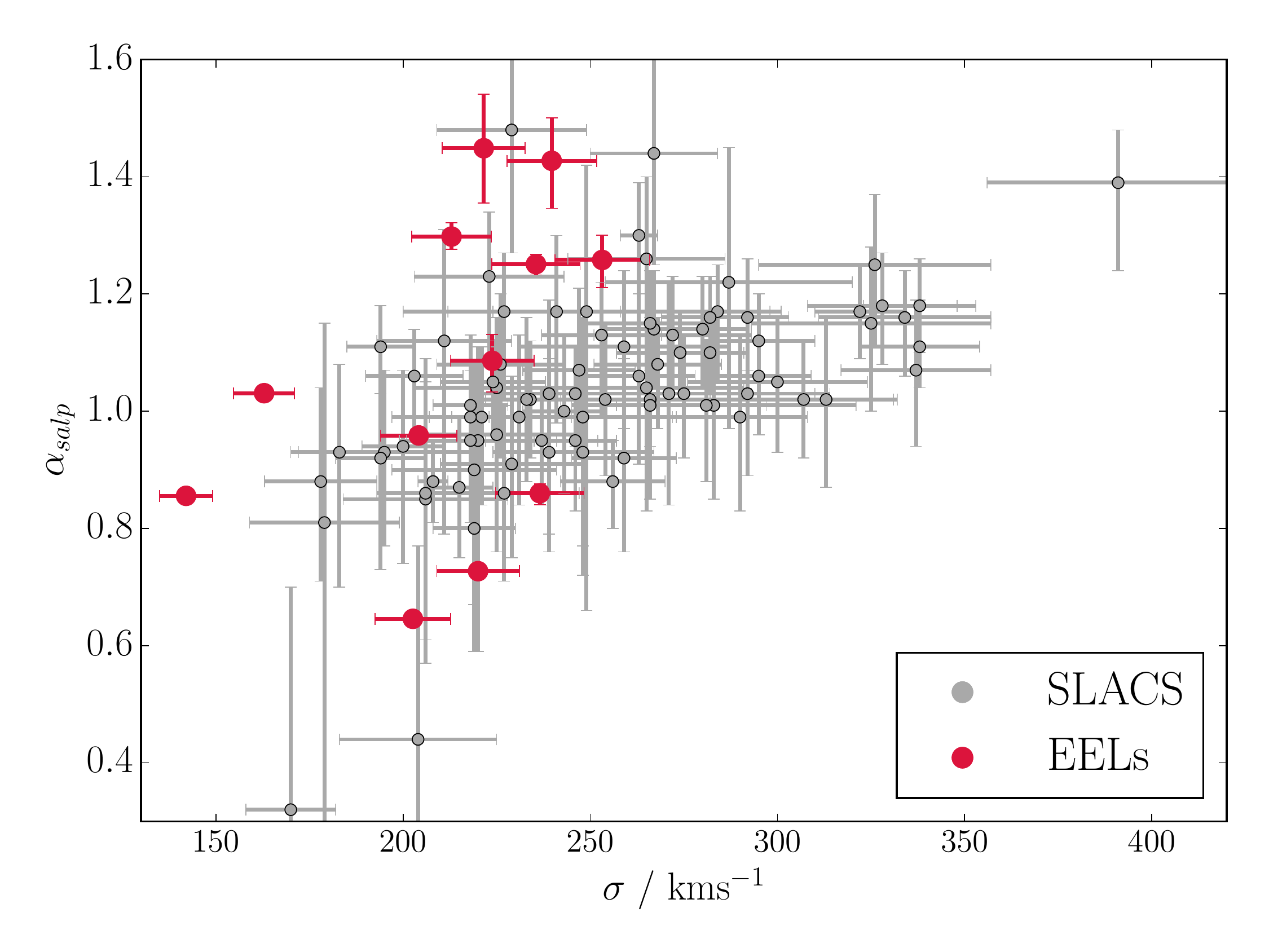}\hfill
\caption{Left: inference on the IMF mismatch parameter $\alpha_c$ and inner halo slope $\gamma(<R_e)$ for the full EELs sample. Whilst all systems have stellar masses requiring IMFs which are at least as heavy as Chabrier, two halo structure seems to form an almost bimodal population, with a small number of systems having sub-NFW haloes ($\gamma(<R_e) \lesssim 0.5$) and the majority having cuspy haloes ($\gamma(<R_e)\gtrsim 1.5$). For two systems, the halo makes a negligible contribution to the lensing, meaning that the halo slope is virtually unconstrained. Right: Variations in IMF mismatch parameter relative to a \emph{Salpeter} IMF $\alpha_{salp}$ as a function of stellar velocity dispersion $\sigma_{\star}$ for the EELs and the SLACS lenses. Our sample is too small to allow an investigation of whether these quantities are correlated, but they seem broadly consistent with SLACS.}
\label{fig:chap8fig3}
\end{figure*}

\begin{figure*}
\centering
\includegraphics[trim=10 10 10 10,clip,width=0.49\textwidth]{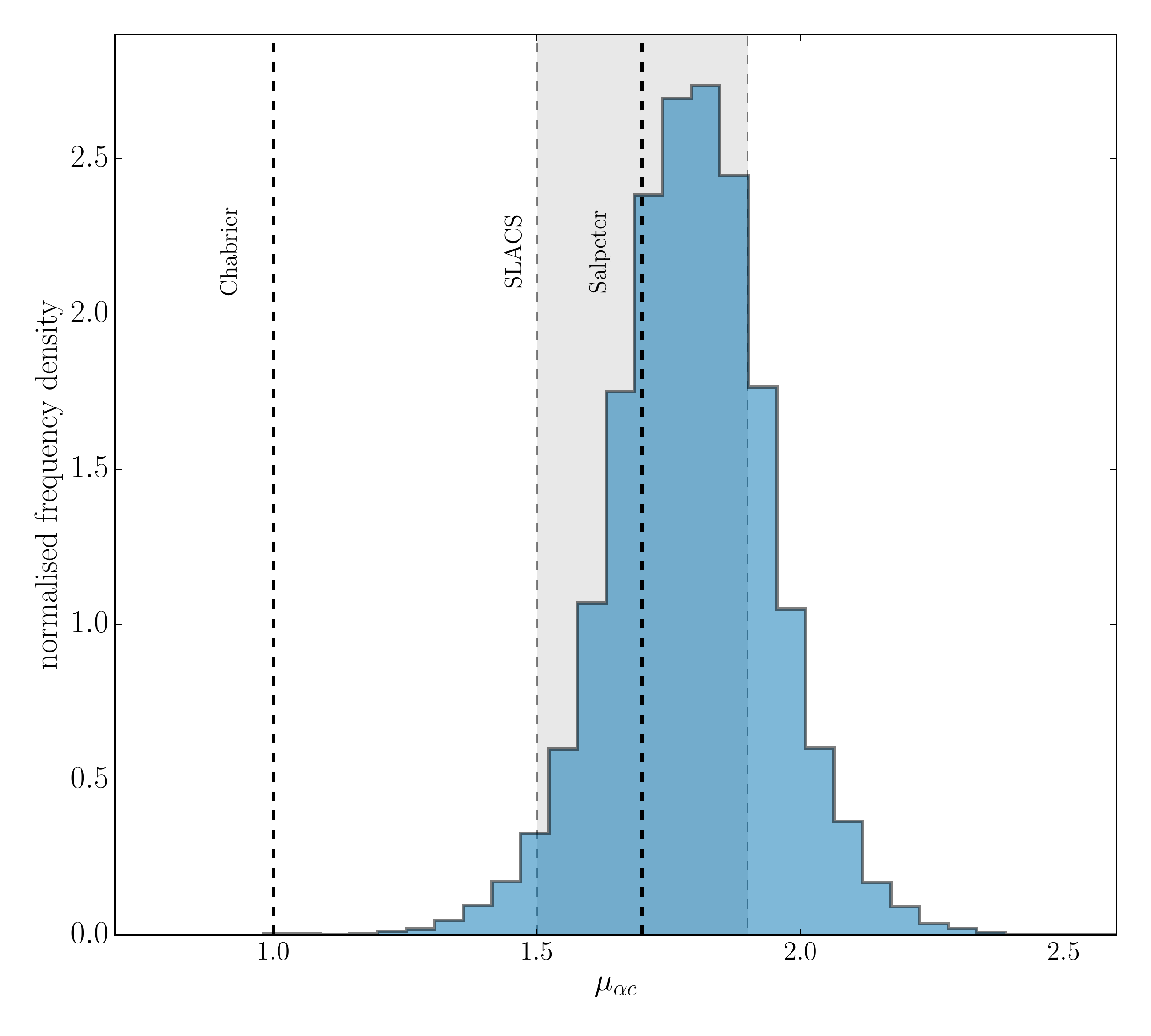}\hfill
\includegraphics[trim=10 10 10 10,clip,width=0.49\textwidth]{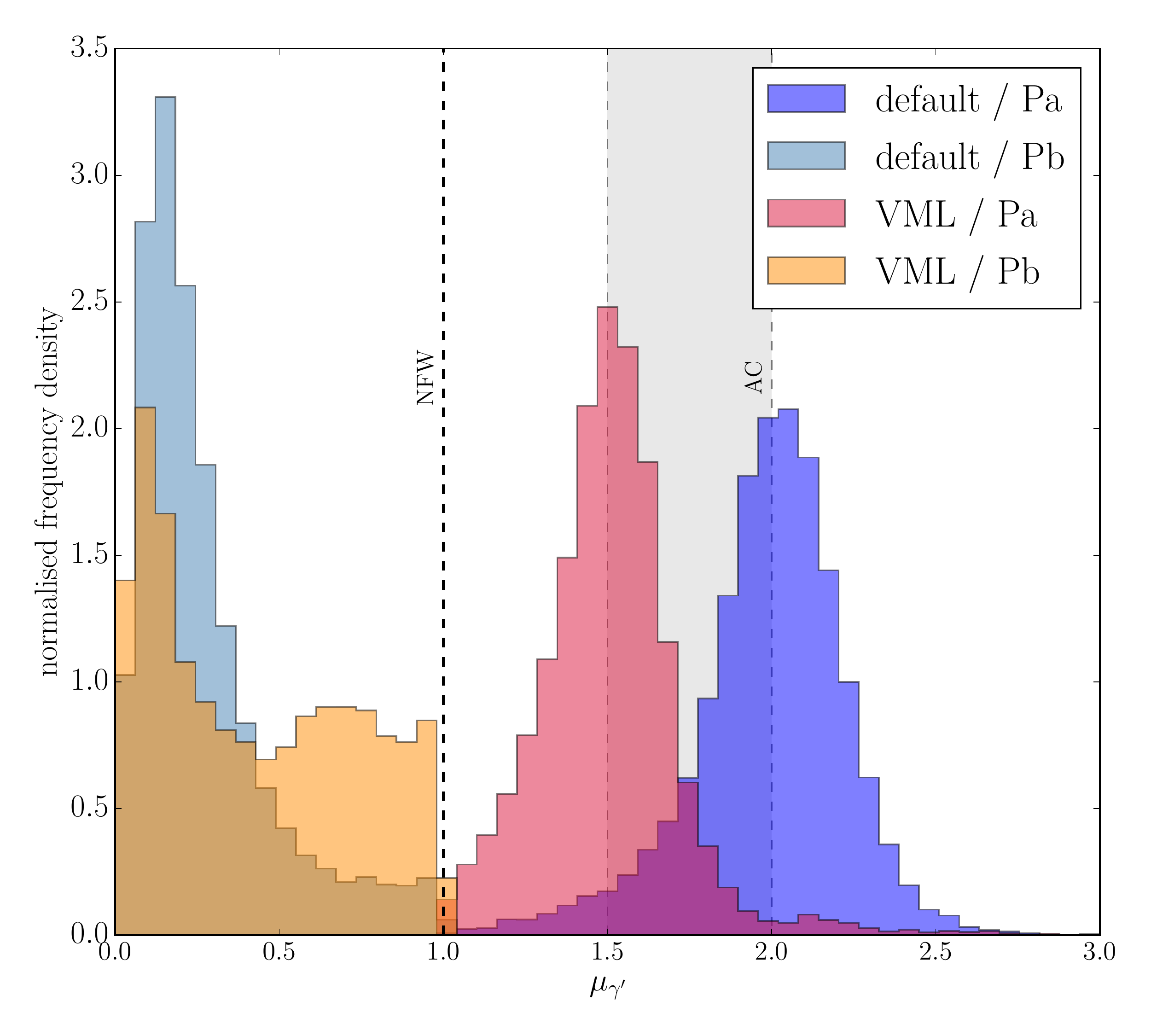}\hfill
\caption{Left: Hierarchical inference on the mean global $\alpha_c$ of the sample, $\mu_{\alpha c}$, as described in Section 3.1, and compared with expectations assuming Chabrier and Salpeter IMFs (dashed lines) and with the results for SLACS from \citet{Auger2010b} (with the width of the latter constraint due to the different halo models that were considered in that study). The EELs have IMF mismatch parameters which are consistent with a Salpeter IMF, but inconsistent with a Chabrier IMF at $>99\%$ confidence. Right: Hierarchical inference on the mean global halo slope within the effective radius, $\mu_{\gamma'}$, as described in Sections 3.2 (default models) and 3.4.1 (VML models) and compared with expectations from dark-matter-only simulations and analytic adiabatic contraction models (dashed lines). When modelled as being drawn from a pair of Gaussians, the EELs clearly separate into two distinct populations with different halo structures. In both sets of models, one of these populations is consistent with expectations from adiabatic contraction models (labelled Population A or Pa in the Figure), whilst the other (Pb) is dominated by much shallower slopes.}
\label{fig:hypers}
\end{figure*}

\begin{table*}
\centering
\begin{tabular}{ccccccccc}\hline
EEL & $\log (M_{DM}/M_{\odot})$ & $r_s$ (kpc) & $\gamma$ & $M_{\star} (\times 10^{11} M_{\odot})$ & $\alpha_c$ & $\sigma_{\star}$ (kms$^{-1}$)& $f_{DM}(<5\mathrm{kpc})$ \\\hline
J0837 & $ 10.27 _{- 0.09 }^{+ 0.09 }$ &  $ 88.9 _{- 61.5 }^{+ 86.4 }$ &  $ 1.69 _{- 0.21 }^{+ 0.15 }$ &  $ 4.07 _{- 0.15 }^{+ 0.14 }$ & $ 2.14 _{- 0.08 }^{+ 0.07 }$ & $ 253.2 \pm 12.7 $ & $0.21 \pm 0.01$\\
J0901 & $ 10.53 _{- 0.05 }^{+ 0.05 }$ &  $ 4.6 _{- 0.8 }^{+ 1.2 }$ &  $ 1.88 _{- 0.07 }^{+ 0.06 }$ &  $ 1.63 _{- 0.09 }^{+ 0.08 }$ & $ 2.66 _{- 0.15 }^{+ 0.14 }$ & $ 239.7 \pm 12.0 $ & $ 0.33 \pm 0.02 $\\
J0913 & $ 10.20 _{- 0.05 }^{+ 0.05 }$ &  $ 156.1 _{- 57.2 }^{+ 40.4 }$ &  $ 0.27 _{- 0.16 }^{+ 0.26 }$ &  $ 2.59 _{- 0.04 }^{+ 0.05 }$ & $ 2.27 _{- 0.04 }^{+ 0.04 }$ & $ 212.9 \pm 10.7 $ & $ 0.29 \pm 0.01$\\
J1125 & $ 10.55 _{- 0.01 }^{+ 0.02 }$ &  $ 187.3 _{- 47.2 }^{+ 30.1 }$ &  $ 0.18 _{- 0.09 }^{+ 0.16 }$ &  $ 5.27 _{- 0.12 }^{+ 0.10 }$ & $ 1.93 _{- 0.04 }^{+ 0.04 }$ & $ 236.5 \pm 11.8 $ & $ 0.50 \pm 0.03 $\\
J1144 & $ 9.97 _{- 0.25 }^{+ 0.18 }$ &  $ 100.3 _{- 73.7 }^{+ 74.5 }$ &  $ 2.08 _{- 0.14 }^{+ 0.10 }$ &  $ 2.32 _{- 0.11 }^{+ 0.10 }$ & $ 1.97 _{- 0.10 }^{+ 0.08 }$ & $ 223.8 \pm 11.2 $ & $0.09 \pm 0.00 $\\
J1218 & $ 10.27 _{- 0.03 }^{+ 0.03 }$ &  $ 3.4 _{- 0.4 }^{+ 0.3 }$ &  $ 0.31 _{- 0.18 }^{+ 0.29 }$ &  $ 3.02 _{- 0.04 }^{+ 0.04 }$ & $ 2.07 _{- 0.03 }^{+ 0.03 }$ & $ 235.5 \pm 11.8 $ & $ 0.31 \pm 0.02 $\\
J1323 & $ 10.64 _{- 0.01 }^{+ 0.01 }$ &  $ 182.6 _{- 5.3 }^{+ 2.6 }$ &  $ 0.06 _{- 0.01 }^{+ 0.01 }$ &  $ 1.74 _{- 0.02 }^{+ 0.02 }$ & $ 1.39 _{- 0.01 }^{+ 0.01 }$ & $ 142.0 \pm 7.1 $ & $ 0.76 \pm 0.04 $\\
J1347 & $ 9.77 _{- 0.10 }^{+ 0.10 }$ &  $ 56.0 _{- 32.7 }^{+ 47.6 }$ &  $ 2.08 _{- 0.23 }^{+ 0.17 }$ &  $ 1.89 _{- 0.03 }^{+ 0.03 }$ & $ 1.15 _{- 0.02 }^{+ 0.02 }$ & $ 202.6 \pm 10.1 $ & $ 0.08 \pm 0.02 $\\
J1446 & $ 7.06 _{- 1.29 }^{+ 1.33 }$ &  $ 99.1 _{- 63.7 }^{+ 59.1 }$ &  $ 1.08 _{- 0.75 }^{+ 1.05 }$ &  $ 1.42 _{- 0.00 }^{+ 0.00 }$ & $ 1.79 _{- 0.00 }^{+ 0.00 }$ & $ 162.9 \pm 8.1 $ & $ 0.00 \pm 0.02 $\\
J1605 & $ 10.48 _{- 0.07 }^{+ 0.06 }$ &  $ 67.3 _{- 40.4 }^{+ 55.9 }$ &  $ 1.71 _{- 0.13 }^{+ 0.09 }$ &  $ 2.13 _{- 0.14 }^{+ 0.14 }$ & $ 1.85 _{- 0.12 }^{+ 0.12 }$ & $ 221.6 \pm 11.1 $ & $ 0.38 \pm 0.02 $\\
J1606 & $ 8.87 _{- 1.24 }^{+ 0.35 }$ &  $ 115.2 _{- 73.2 }^{+ 63.6 }$ &  $ 0.82 _{- 0.56 }^{+ 0.88 }$ &  $ 2.77 _{- 0.03 }^{+ 0.02 }$ & $ 1.27 _{- 0.01 }^{+ 0.01 }$ & $ 220.0 \pm 11.0 $ & $ 0.03 \pm 0.02 $\\
J2228 & $ 9.62 _{- 0.07 }^{+ 0.06 }$ &  $ 0.9_{- 0.1 }^{+ 0.1 }$ &  $ 2.60 _{- 0.13 }^{+ 0.12 }$ &  $ 2.89 _{- 0.02 }^{+ 0.02 }$ & $ 1.67 _{- 0.01 }^{+ 0.01 }$ & $ 204.2 \pm 10.2 $ & $ 0.04 \pm 0.02 $\\\hline
\end{tabular}
\caption{Default models for all the EELs. From left to right, the columns show the system name, the projected dark matter mass within a 2.5kpc aperture $\log(M_{DM}/M_{\odot})$, the halo scale radius $r_s$, the halo inner slope $\gamma$, total stellar mass $M_{\star}$, IMF mismatch parameter relative to a Chabrier IMF $\alpha_{c}$, stellar velocity dispersion $\sigma_{\star}$, and the projected dark matter fraction within a 5 kpc aperture $f_{DM}(<$5kpc); uncertainties are statistical only. In the context of this model, most systems have high stellar masses and steep halo slopes (see Table~\ref{tab:chap8tab3}).}
\label{tab:chap8tab2}
\end{table*}

\begin{table*}
\centering
\begin{tabular}{ccccccc}\hline
model & $X$ & $\mu_{X1}$ & $\tau_{X1}$ & $\mu_{X2}$ & $\tau_{X2}$ & $f_1$ \\\toprule
default & $\alpha_c$ & $1.80 \pm 0.14$ & $0.49_{-0.10}^{+0.14}$ & -- & -- & -- \\
default & $\gamma(R<R_e)$\phantom{$\dagger$} & $1.41 \pm 0.30$ & $0.98_{-0.19}^{+0.28}$ & -- & -- & --\\
default & $\gamma(R<R_e)$$\dagger$ & $2.01_{-0.22}^{+0.19}$ & $0.45_{-0.13}^{+0.25}$ & $0.10_{-0.10}^{+0.33}$ & $0.20_{-0.11}^{+0.34}$ & $0.61_{-0.15}^{+0.14}$ \\\midrule

VML & $\alpha_c$ & $1.64 \pm 0.18$ & $0.64_{-0.13}^{+0.19}$ & -- & -- & -- \\
VML & $\gamma(R<R_e)$\phantom{$\dagger$} & $1.12 \pm 0.22$ & $0.68_{-0.13}^{+0.20}$ & -- & -- & --\\
VML & $\gamma(R<R_e)$$\dagger$ & $1.50_{-0.19}^{+0.16}$ & $0.33_{-0.14}^{+0.25}$ & $0.09_{-0.09}^{+0.70}$ & $0.48_{-0.36}^{+0.31}$ & $0.65_{-0.26}^{+0.17}$ \\
VML & $\mu_{ML}$ & $0.02 \pm 0.07$ & $0.22_{-0.04}^{+0.06}$ & -- & -- & -- \\\bottomrule

\end{tabular}
\caption[Hierarchical models of the EELs as a population]{We characterise the EELs lens sample using a hierarchical model in which the property $X$ of each EEL is drawn from either one or two Gaussian parent distributions $X \sim N(\mu,\tau^2)$. Where two Gaussians are used, $f_1$ represents the relative weight of the first Gaussian. In general, the EELs lenses have stellar masses consistent with Salpeter IMFs, dark halo slopes consistent with simple adiabatic contraction models. $\dagger$ Modelling the EELs as being drawn from two Gaussians, rather than one.}
\label{tab:chap8tab3}
\end{table*}

\subsection{Dark matter}
\label{sec:chap8sec3sub1sub2}

For two lenses (J1446 and J1606), the halo makes a negligible contribution to the Einstein mass, such that the halo properties are virtually unconstrained (Figure~\ref{fig:chap8fig3}, left). The remaining lenses appear to form a virtually bimodal distribution, with inner slopes which are either very cusped ($\gamma \gtrsim 1.5$) or cored ($\gamma \lesssim 0.5$). To make a meaningful comparison between systems with different scale radii and inner slopes, we calculate the mass-weighted slope within the effective radius as
\begin{equation}
\begin{split}
&~~~~~\gamma' = \gamma (r<R_e) \\
= -\frac{1}{M_{DM}(r)}&\int_0^{R_e} 4 \pi r^2 \rho_{DM}(r)\frac{d\log \rho_{DM}}{d \log r} dr \\
&= 3 - \frac{4 \pi R_e^3\rho_{DM}(R_e)}{M(R_e)}
\end{split}
\end{equation}
for dark halo 3D mass $M_{DM}(r)$ and density $\rho_{DM}(r)$ \citep[e.g.][]{Newman2015}. We then construct a hypermodel as in Section~\ref{sec:chap8sec3sub1} in which the $\gamma'$ distributions of the EELs are drawn from a Gaussian distribution $\gamma' \sim N(\mu_{\gamma'},\tau_{\gamma'}^2)$. In this model, we find $\mu_{\gamma} = 1.41 \pm 0.30$, $\tau_{\gamma'} = 0.98_{-0.19}^{+0.28}$, which is consistent with expectations from both dark-matter-only simulations (shown as the vertical line marked `NFW' in Figure 4) and the adiabatic contraction (`AC') models of \citet{Gnedin2004}. However, this does not describe the near-bimodal distribution of $\gamma'$ that is evident in Figure 3 (left). We therefore increase the complexity of our hypermodel such that the distribution is composed of two Gaussians with different means and dispersions, i.e. $P(D|\omega)$ in Equation 7 becomes
\begin{equation}
\centering
P(D|\omega) = \prod_i \int d\gamma' P_i(D_i|\gamma_i) P(\gamma';\omega)
\end{equation}
where
\begin{equation}
\centering
P(\gamma',\omega) = f_1 N(\mu_{\gamma' 1},\tau_{\gamma' 1}^2) + (1 - f_1) N(\mu_{\gamma' 2},\tau_{\gamma' 2}^2)
\end{equation}
and the hyperparameters are $\omega = (\mu_{\gamma' 1}, \tau_{\gamma' 1}, \mu_{\gamma' 2}, \tau_{\gamma' 2}, f_1)$, such that $f_1$ describes the relative weighting of the two Gaussians. In this case, we find $\mu_{\gamma' 1} = 2.01_{-0.22}^{+0.19}$, $\mu_{\gamma' 2} = 0.10_{-0.10}^{+0.33}$ and $f_1 = 0.61_{-0.15}^{+0.14}$ (see Table 3 and Figure 4, right). We discuss possible physical explanations for these results in Section 4.2.

\subsection{Total mass}

Previous studies have focused on constraining the profile of the \emph{total} mass density, as opposed to that of the separate components (see Figure 5), either by combining aperture mass measurements in the form of the Einstein radius $R_E$ and the velocity dispersion \citep[e.g.][]{Sonnenfeld2013} or directly from pixel-based lens modelling \citep[e.g.][]{Barnabe2011}. These studies have generally found that, when the total mass density is modelled with a power law, the power-law index $\gamma_{pl}$ is close to isothermal \citep[e.g.][]{Treu2002,Koopmans2003,Koopmans2006,Koopmans2009}. However, since the total mass density is not scale-free in reality, our ability to interpret $\gamma_{pl}$ physically has remained limited. To first order, $\gamma_{pl}$ should represent an approximation to the average density slope over the radial range in which mass probes exist (e.g. the region of the Einstein arcs, or between the Einstein arcs and the aperture in which the velocity dispersion is measured). In the case of aperture mass measurements, \citet{Sonnenfeld2013} investigated this assumption more quantitatively and showed that $\gamma_{pl}$ agrees more closely with the mass-weighted mean density slope within $\sim 0.5-1 R_{e}$ (with $R_E/R_e$ of order 1) than with the local logarithmic density slope at the Einstein radius, though the agreement is only approximate. Since we have modelled the EELs at the pixel level using parameterisations for both the total mass \citep{Oldham2017a} and the dark and luminous mass components separately (this work), we are now in a position to carry out a similar test for the case of pixel-based lens modelling. Given that modelling the total mass with a power law is computationally cheaper than performing the decomposition using realistic profiles for the individual mass components, a better understanding of the meaning of $\gamma_{pl}$ will be valuable for future studies.

We update the \citet{Oldham2017a} models to incorporate the kinematic information, and show a comparison of the inferred projected mass profiles in the total mass (TM) and composite mass (CM) cases in Figure 5. Whilst the Einstein mass is robustly reproduced by both sets of models, the \emph{slope} of the total profile at the Einstein radius clearly differs in each case. Similarly to \citet{Sonnenfeld2013}, then, this strongly disfavours the view that $\gamma_{pl}$ is tracing the local logarithmic density slope at the Einstein radius. To investigate this further, we use our CM models to calculate 
\begin{enumerate}
\item the mass-weighted slope within the Einstein radius, 
\begin{equation}
\gamma' = \gamma(r<R_E)
\end{equation}
according to Equation 8;
\item the local logarithmic slope at the Einstein radius, 
\begin{equation}
\gamma_{pl}(R_E) = - \frac{\mathrm{d}\ln \rho}{\mathrm{d}\ln r}\Bigm\lvert_{R_E}
\end{equation}
\item the average slope (or power-law interpolation) between two radii $(r_1, r_2)$, 
\begin{equation}
\gamma_{avg}^{(r_1,r_2)} = \frac{\ln[\rho(r_2)/\rho(r_1)]}{\ln (r_1/r_2)}.
\end{equation}
\end{enumerate}
These calculations are summarised in Table 4. We find, again similarly to \citet{Sonnenfeld2013}, that the slope estimators which account for the averaging of the slope within radii $\lesssim R_E$ provide the best agreement with $\gamma_{pl}$ as measured in the TM models. However, we also stress that the correspondence is only approximate, and that inference on the mass distribution is also dependent on the properties of the source and the lensing configuration.

\begin{figure*}
\centering
\includegraphics[trim=20 40 20 20,clip,width=\textwidth]{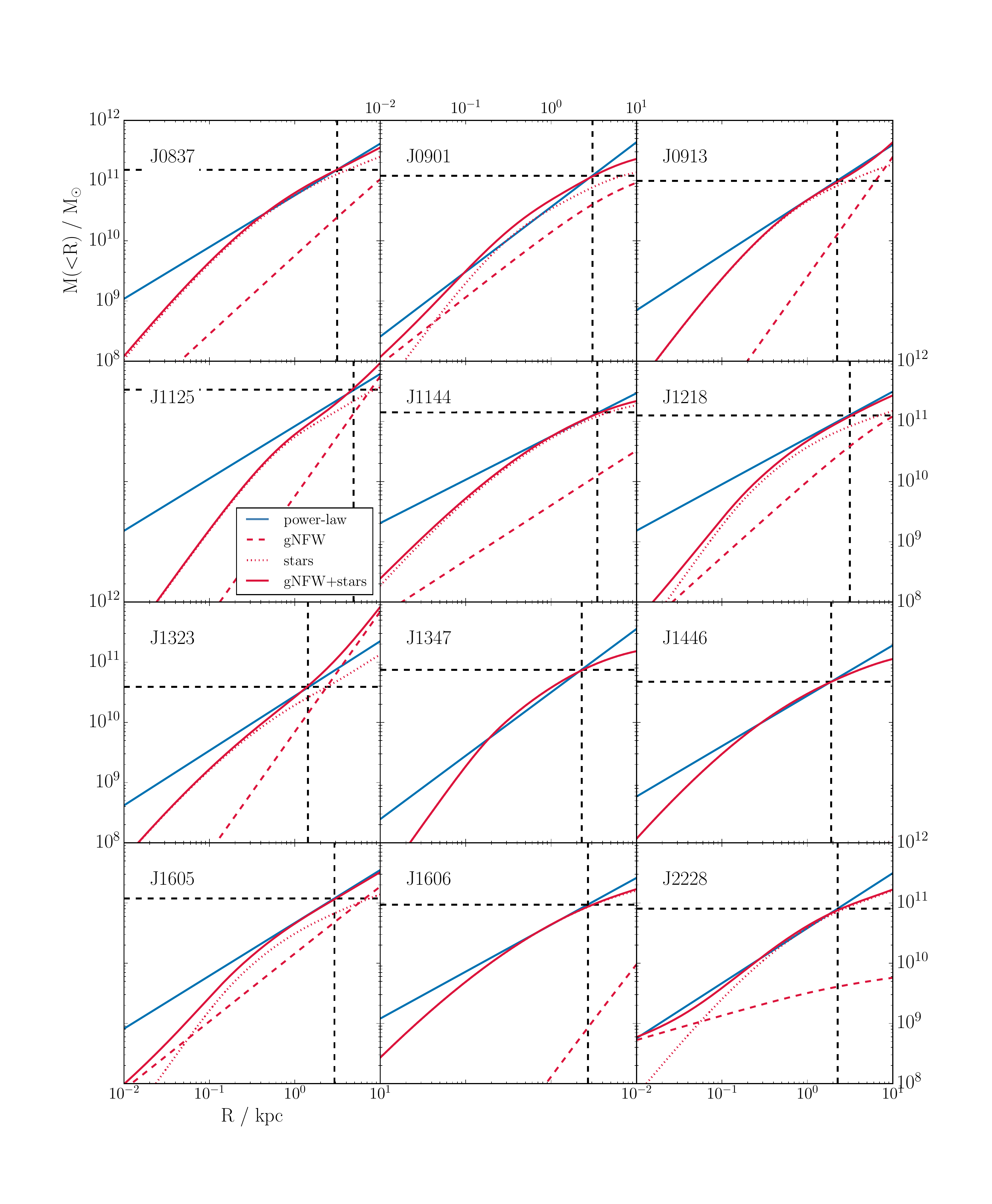}\hfill
\caption{Comparison of the cumulative projected mass for models treating the total mass as a power law (blue) and separating the dark and luminous components (red, with dark and luminous mass shown by dashed and dotted lines respectively). The Einstein radius and Einstein mass are marked with black dashed lines. Though all models robustly reproduce the Einstein mass, the composite models do not resemble power laws at the Einstein radius for all EELs, and the local logarithmic slope does not necessarily agree between the two models. As shown in Table 4, it is approximately the average slope within the Einstein radius that the total-mass models robustly recover. }
\end{figure*}

\begin{table*}
\centering
\begin{tabular}{cccccc}\hline
EEL & $\gamma_{pl}$ & $\gamma(R=R_{E})$ & $\gamma_{avg}^{(0.5,1.5R_E)}$ & $\gamma_{avg}^{(0.05,1.5R_E)}$ & $\gamma'(R<R_E)$   \\\hline
J0837 & $ 2.14 \pm 0.01 $ & $ 2.42 \pm 0.12 $ & $ 2.42 \pm 0.12 $ & $ 2.17 \pm 0.11 $ & $ 2.27 \pm 0.11 $ \\
J0901 & $ 1.92 \pm 0.02 $ & $ 2.32 \pm 0.12 $ & $ 2.29 \pm 0.11 $ & $ 2.11 \pm 0.11 $ & $ 2.15 \pm 0.11 $ \\
J0913 & $ 2.08 \pm 0.02 $ & $ 2.69 \pm 0.13 $ & $ 2.62 \pm 0.13 $ & $ 2.03 \pm 0.10 $ & $ 2.29 \pm 0.11 $ \\
J1125 & $ 2.13 \pm 0.01 $ & $ 1.91 \pm 0.10 $ & $ 1.89 \pm 0.09 $ & $ 1.90 \pm 0.10 $ & $ 2.00 \pm 0.10 $ \\
J1144 & $ 2.28 \pm 0.02 $ & $ 2.59 \pm 0.13 $ & $ 2.58 \pm 0.13 $ & $ 2.29 \pm 0.11 $ & $ 2.33 \pm 0.12 $ \\
J1218 & $ 2.23 \pm 0.01 $ & $ 2.32 \pm 0.12 $ & $ 2.33 \pm 0.12 $ & $ 2.08 \pm 0.10 $ & $ 2.19 \pm 0.11 $ \\
J1323 & $ 2.09 \pm 0.03 $ & $ 2.39 \pm 0.12 $ & $ 2.34 \pm 0.12 $ & $ 2.03 \pm 0.10 $ & $ 2.15 \pm 0.11 $ \\
J1347 & $ 1.95 \pm 0.02 $ & $ 2.90 \pm 0.15 $ & $ 2.87 \pm 0.14 $ & $ 2.62 \pm 0.13 $ & $ 2.56 \pm 0.13 $ \\
J1446 & $ 2.16 \pm 0.02 $ & $ 2.51 \pm 0.13 $ & $ 2.48 \pm 0.12 $ & $ 2.20 \pm 0.11 $ & $ 2.27 \pm 0.11 $ \\
J1605 & $ 2.12 \pm 0.02 $ & $ 2.22 \pm 0.11 $ & $ 2.25 \pm 0.11 $ & $ 2.06 \pm 0.10 $ & $ 2.18 \pm 0.11 $ \\
J1606 & $ 2.18 \pm 0.10 $ & $ 2.60 \pm 0.13 $ & $ 2.56 \pm 0.13 $ & $ 2.30 \pm 0.11 $ & $ 2.32 \pm 0.12 $ \\
J2228 & $ 2.09 \pm 0.03 $ & $ 2.71 \pm 0.14 $ & $ 2.63 \pm 0.13 $ & $ 2.16 \pm 0.11 $ & $ 2.27 \pm 0.11 $ \\\hline
\end{tabular}
\caption{Comparison of lens models in which a power law profile with slope $\gamma_{pl}$ is assumed for the total mass with slope estimators from the composite models. As presented in Section 3.3, we consider the local logarithmic slope $\gamma(R = R_E)$, the average slope or power-law interpolation between two radii $R_1$ and $R_2$, $\gamma_{avg}^{(R_1,R_2)}$, and the mean mass-weighted slope within $R_E$, $\gamma'(R<R_E)$. In general, the inferred slopes $\gamma_{pl}$ agree most closely with the average slope within $\sim R_E$. On the other hand, the local logarithmic slope $\gamma(R=R_E)$ does not provide an accurate match to $\gamma_{pl}$. We emphasise, however, that the consistency remains approximate.}
\end{table*}

\subsection{Alternative models}

Under the assumptions of our default model, we find the EELs to have Salpeter-like stellar-mass-to-light ratios and halo structures which are inconsistent with dark-matter-only predictions, implying both that these ETGs formed their stars in different physical conditions from the Milky Way and that baryonic processes play an important role in determining the central dark matter distribution. However, it is possible that our modelling assumptions may be biasing our inference on $\Upsilon_{\star}$ and $\gamma(r<R_e)$ (with the latter defined by Equation 8) -- in particular, the stellar-mass-to-light ratio may not be spatially uniform, and the stellar orbits may not be isotropic. Here, we investigate the nature of these potential biases and the extent to which they can be overcome or quantified. We present a further suite of systematic tests in Appendix A.

\subsubsection{Models with a spatially varying $\Upsilon_{\star}$}
\label{sec:chap8sec3sub2}

A possible explanation for the cuspy halo structure and high IMF mismatch parameters that we infer for the majority of the EELs may be the presence of negative radial gradients in their stellar-mass-to-light ratios, due to trends in stellar population properties such as the age, metallicity and the IMF \citep[e.g.][]{MartinNavarro2015,vanDokkum2016,M873}. As our default model does not allow for such a gradient, it may, in some cases, be forcing the halo profile to be steep in order to reproduce the slope of the \emph{total} mass profile. Equally, it may be driving our inference to high stellar mass-to-light ratios which may only really exist in the central regions, where the surface brightness profile is best constrained. Since strong lensing and dynamics are purely gravitational probes, we cannot formally distinguish between gradients in $\Upsilon_{\star}(R)$ and $\rho_{DM}$; however, by assuming simple parameterisations for each, we can make some steps to explore the degeneracies between them. We stress, however, that our results in this scenario are dependent on the parametric forms that we adopt. With this limitation in mind, we construct models for the EELs in which the stellar-mass-to-light ratio is given by
\begin{equation}
\log \Upsilon_{\star} = \log\Upsilon_{\star,1} - \mu_{ML} \log R
\end{equation}
where $\Upsilon_{\star,1}$ is the stellar mass-to-light ratio at a radius of 1 kpc. We make the deflection angle computations tractable by fitting the S\'ersic profiles with multi-Gaussian expansions (MGEs) using the publicly available code of \citet{Cappellari2002}, and precalculating the deflections for each MGE component individually; these can then be scaled by the value of $\Upsilon_{\star}(R)$ at the width $\sigma_k$ of the $k^{th}$ Gaussian. We note that, although the limiting behaviour of this model is unrealistic, our data only probe the central regions of each galaxy, which makes its large-radius behaviour unimportant. Equally, due to the MGE decomposition, the finite width of the innermost Gaussian prevents $\Upsilon_{\star}(R\to0)$ from becoming infinitely large\footnote{Specifically, $\Upsilon_{\star}(R=0) = \Upsilon_{\star,1}\sigma_1^{-\mu_{ML}}$ where $\sigma_1$ is the width of the innermost Gaussian component in kpc.}. We use a uniform prior on the stellar-mass-to-light ratio gradient $-1 \leq \mu_{ML} \leq 1$ to reflect our lack of prior knowledge of the gradient of $\Upsilon_{\star}(R)$. We refer to these as the \emph{v}arying stellar-\emph{m}ass-to-\emph{l}ight ratio models, or VML models, in what follows.

\begin{figure}
\centering
\includegraphics[trim = 20 100 20 170,clip,width=0.48\textwidth]{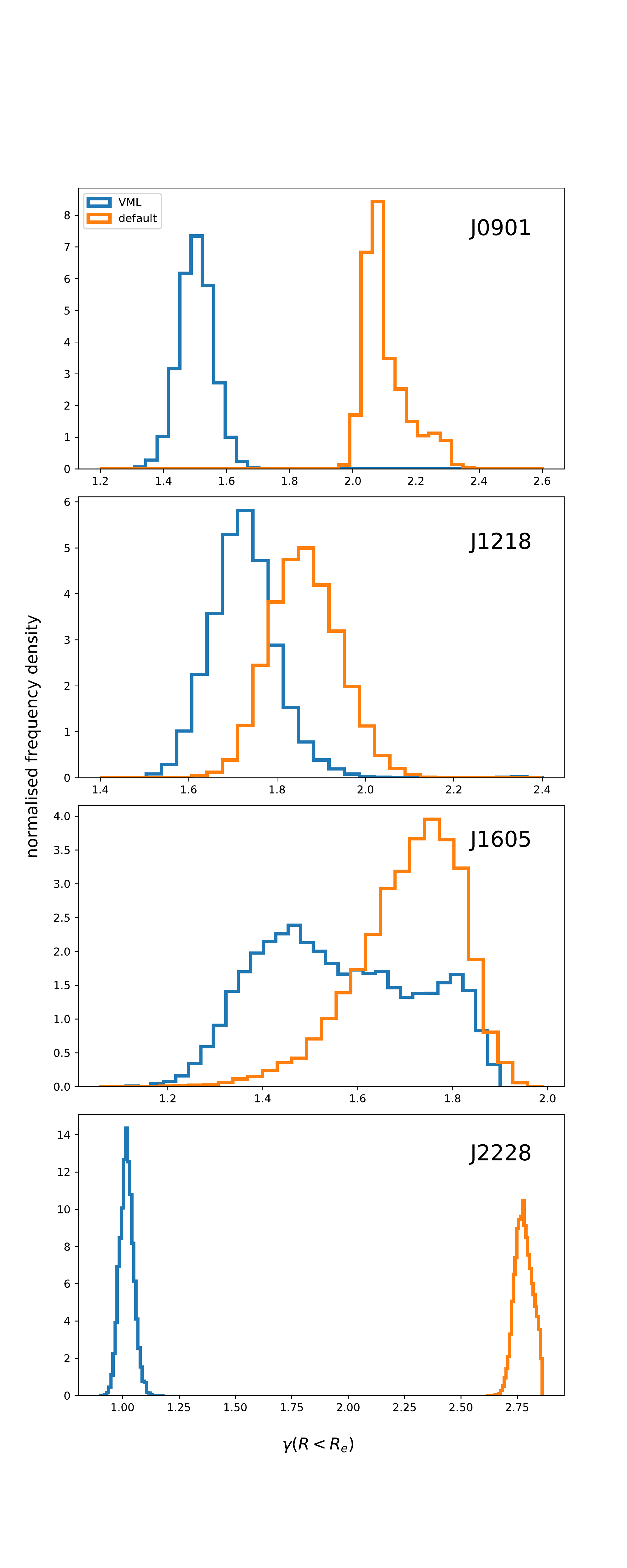}\hfill
\caption{Shifts in the inference on the mass-weighted halo slope within the effective radius, $\gamma(R<R_e)$, for the four systems which strongly prefer non-zero radial stellar-mass-to-light ratio gradients (i.e. $|\mu_{ML}|>0.1$). For three of these systems (J0901, J1218, J2228), radially declining stellar-mass-to-light ratios are preferred, resulting in less cuspy halo profiles. For J2228, this effect is particularly striking, and offers a solution to the problem of this system's default model in which the halo profile is centrally steeper than that of the stellar mass. For the fourth system (J1605), a radially increasing stellar-mass-to-light ratio is preferred, but the effect of this shift on the halo profile is weaker. }
\label{fig:gammaShifts}
\end{figure}

Our VML and default models are nested models, such that the VML model becomes the default when $\mu_{ML} = 0$. Indeed, for eight out of the twelve systems, we infer $|\mu_{ML}| < 0.1$ which, given the systematic uncertainty due to our imposed parametric profile, we take to be consistent with a spatially uniform stellar-mass-to-light ratio. Of the remaining four systems, three (J0901, J1218, J2228) prefer radially declining stellar-mass-to-light ratios ($\mu_{ML} > 0.1$). We therefore find them to also have less cuspy haloes than in their default models (Figure~\ref{fig:gammaShifts}). In particular, the inner slope of J2228's halo decreases from  $\gamma = 2.60_{-0.13}^{+0.12}$ to $\gamma = 0.73_{-0.07}^{+0.08}$ due to its strong declining gradient $\mu_{ML} = 0.41 \pm 0.03$, such that the halo density no longer becomes steeper than the stellar mass in the central regions. The final system (J1605) prefers a radially increasing stellar-mass-to-light ratio and its halo structure is modified only weakly compared to its default model. Whilst varying stellar-mass-to-light ratios can be explained by a scenario in which the age, metallicity or IMF of the stellar populations varies spatially, previous studies indicate that these effects should cause the stellar-mass-to-light ratio to rise in the central regions \citep[e.g.][]{MartinNavarro2015}. That our inference on J1605 is counter to these expectations may indicate variety in ETG stellar mass structure, but may equally reflect the limitations of our VML models.

\begin{figure}
\centering
\includegraphics[trim=10 10 10 10,clip,width=0.49\textwidth]{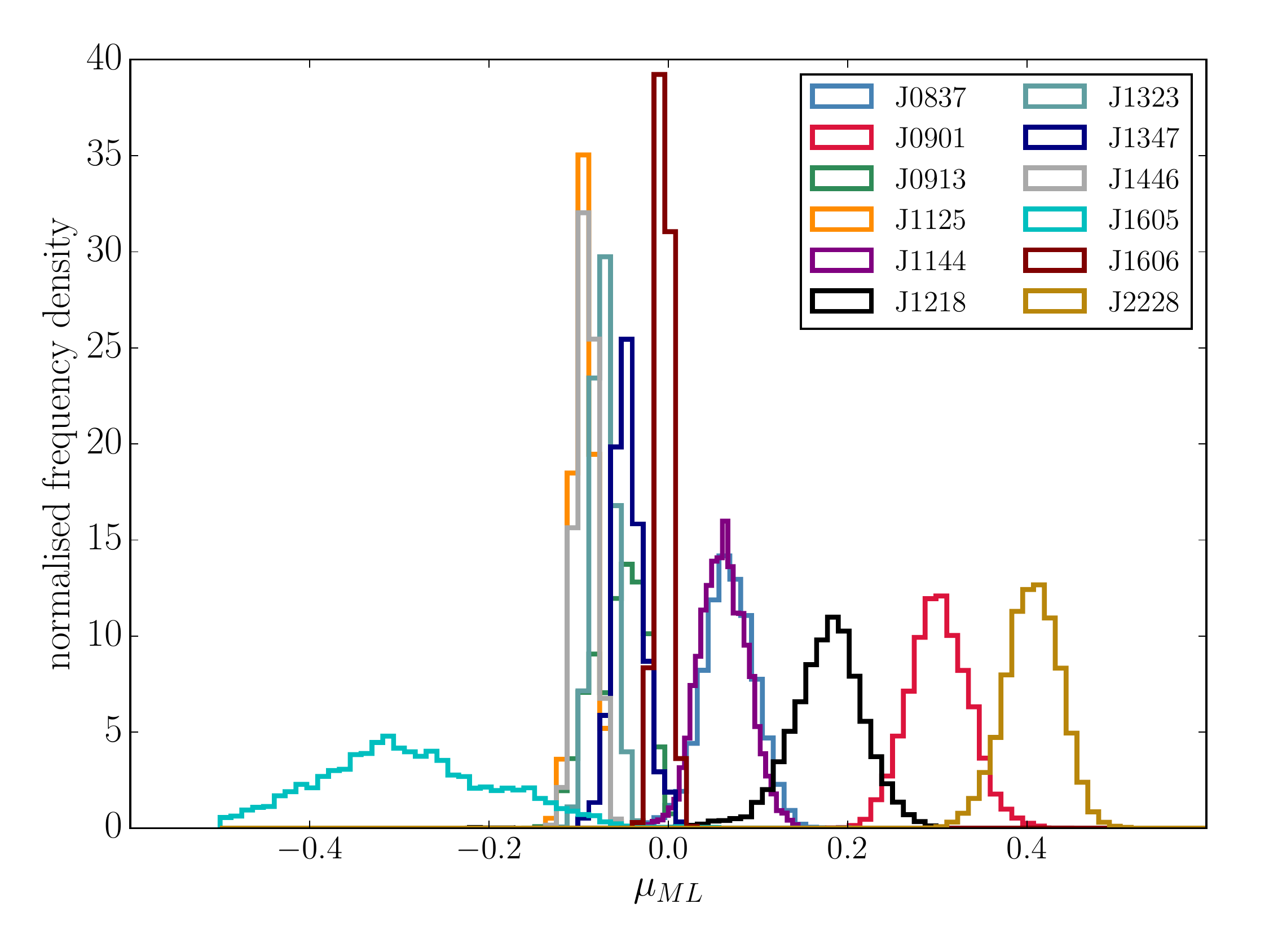}\hfill
\caption{Inference on the gradient of the stellar-mass-to-light ratio, $\mu_{ML}$, for the sample, as defined in Equation 11. For most EELs, $|\mu_{ML}| < 0.1$ within $2\sigma$, but a small number of systems strongly prefer radially declining gradients. For these systems, the negative gradient generally reduces the total stellar mass and the inner halo slope that are inferred. One system (J1605) strongly prefers an increasing radial gradient; this may be due to shortcomings in the parameterisation of $\Upsilon_{\star}(R)$ or in the S\'ersic profiles describing the surface brightness, or it may be real, but we do not investigate it further here.}
\label{fig:muML}
\end{figure}

For the population as a whole, our inferences on $\mu_{\gamma}$ and $\mu_{\alpha_c}$ do not change significantly: we now find $\mu_{\alpha c} = 1.64 \pm 0.18$ and $\mu_{\gamma'} = 1.12 \pm 0.22$ when all systems are modelled as one population, and $\mu_{\gamma' 1} = 1.50_{-0.19}^{+0.16}$, $\mu_{\gamma' 2} = 0.09_{-0.09}^{+0.70}$ when modelled as being drawn from two populations; the latter results are included in Figure 4 (right). Whilst these results are consistent with the default model results within $2\sigma$, we note that the marginally lower value of $\mu_{\gamma,1}$ requires less extreme contraction than the default models. Finally, Figure~\ref{fig:muML} shows our inferences on $\mu_{ML}$ in this paradigm; we find that $\mu_{\mu_{ml}} = -0.02 \pm 0.07$, implying no overall trend across the population. (Note that removing the outlier, J1605, does not change the significance of this result.)

On the one hand, then, this experiment shows that it is possible to reduce the high values of $\gamma'$ that we infer for some lenses. However, the additional flexibility of the VML model does not change the general need for cuspy haloes and Salpeter-like IMFs. Even if the stellar-mass-to-light ratios in these systems do vary radially, the inferred deviations from NFW-like haloes and Chabrier-like IMFs are robust. We also stress that our inferences here may be dependent on our choice of parameterisation for $\Upsilon_{\star}(R)$, and that thorough systematic tests of this dependence are needed before any strong claims can be made about the existence of $\Upsilon_{\star}$ gradients in the EELs.

\subsubsection{Orbital anisotropy}

It is also possible that the assumptions made in our dynamical modelling may bias our inference on the halo slope. Indeed, \citet{Xu2016} have shown that ETGs in the Illustris simulation, with similar stellar velocity dispersions to the EELs, have mildly radial anisotropies $\beta_{\star}(<2R_e) \sim 0.2$, and that modelling their kinematics assuming $\beta_{\star} = 0$ can lead to a systematic overestimation of their \emph{total} density slopes of order $\Delta \gamma_{pl} \sim 0.1$. We explore the possibility that a more flexible dynamical model would reduce the gradient that we require for some of the EELs' \emph{dark matter haloes} by constructing models in which the stellar orbital anisotropy $\beta_{\star}$ is also a free parameter, on which we include a Gaussian prior such that $\beta_{\star} \sim \mathrm{N}(0,0.3^2)$ to encode our expectation, based on previous observational results, that the stellar kinematics at the centres of massive ETGs are close to isotropic \citep[e.g.][]{Cappellari2007}. However, we find that we are unable to constrain the anisotropy based on our data and that whilst, in some cases, we simply recover the prior, in others the anisotropy tends to a very large value, suggesting that it is overfitting the data (which is unsurprising given that our kinematic constraints are weak compared to the lensing information). On the other hand, we find that fixing $\beta_{\star}$ to a mildly radial value can reduce our inference by $\Delta \gamma \sim 0.1$ in some systems. We therefore regard the effect of neglecting the anisotropy as non-negligible, but not a dominating factor in our inference. We emphasise that more detailed kinematic observations -- and dynamical modelling -- of strong lensing systems will be key to overcoming this source of systematic uncertainty in the future.

\subsection{Comparison with simpler models}

Our lens modelling paradigm makes use of a large number of individual image pixels in multiple HST filters to inform the mass model, and therefore has significantly more constraining power than simpler methods in which only robust aperture mass measurements -- i.e. the Einstein radius or mass and the stellar velocity dispersion -- are used as constraints \citep[e.g.][]{Auger2010b,Sonnenfeld2015}. We can therefore use this simpler, more established method to verify the robustness of our results.

We remodel the EELs using aperture mass measurements only. For each EEL, we use the Einstein radius, taken from \citet{Oldham2017a}, to estimate the Einstein mass, and combine this with the stellar velocity dispersion to constrain the same default mass model as in Section 2.2. Note that, though this clearly has many fewer constraints (the two aperture mass measurements as opposed to all the HST pixels), it also has many fewer free parameters since there is now no need to infer the source structure. (Effectively, that was constrained at the same time as the Einstein radius, but the Einstein radius is very robust against changes to a smooth source model. For a parametric source, this removes six non-linear and one linear parameter per S\'ersic component.) Our model therefore consists of parameters $\vec{M} = (M_{\star}, M_{DM}(R<2.5\mathrm{kpc}),\gamma,r_s)$. We show a comparison of our inferences in this case with those obtained using our default model in Figure 8; in all cases, we are consistent within the 2D 2$\sigma$ contours, and the uncertainties are many times smaller when the full pixel data are used. In general, the upper limit on the inference on $M_{\star}$ in the aperture mass models is set by the Einstein radius, and corresponds to the case in which all the lensing is due to the stellar mass; as no dark matter is then required, the contours in the $\gamma$ direction in these regions are unconstrained. It is also interesting to note the nature of the degeneracy in the aperture mass models: the data prefer either high stellar masses or cuspy haloes, neither of which corresponds to the `vanilla' Chabrier IMF + NFW halo model that might be the zeroth order expectation. 

\begin{figure*}
\centering
\includegraphics[trim=20 20 20 20,clip,width=\textwidth]{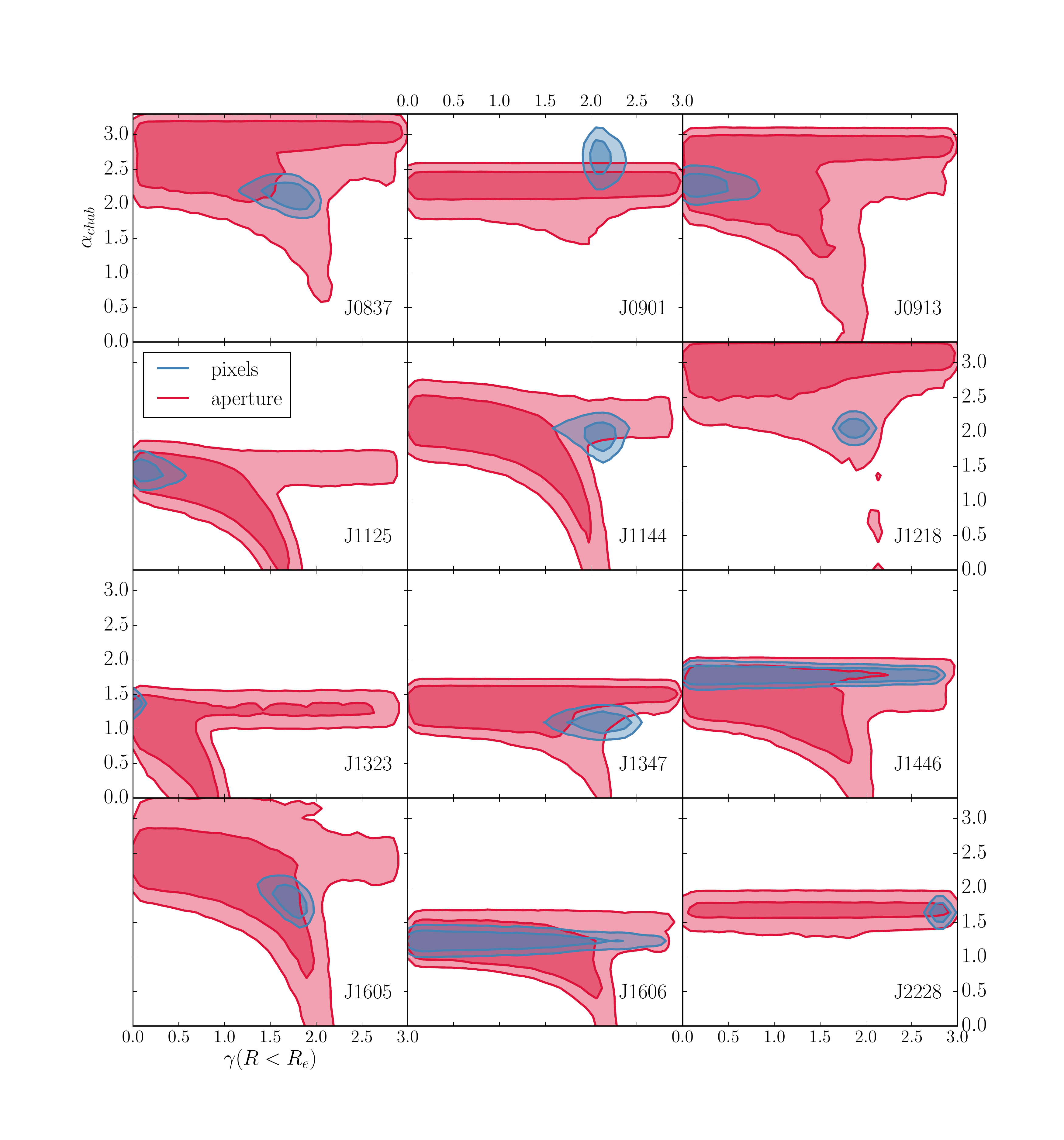}\hfill
\caption{Comparison of our inference on the mean halo slope within $R_e$, $\gamma(<R_e)$, and IMF mismatch parameter $\alpha_c$ in the case of pixel-based models (the main subject of the paper, and shown in blue contours) and simpler models using only aperture mass measurements as constraints (red contours). Our inferences are consistent for all EELs, though the uncertainties are many times smaller when the full pixel data are used due to the much larger number of constraints. For the aperture mass models, the upper limit on the $\alpha_c$ contours is driven by the Einstein mass and corresponds to the case in which the lensing is entirely due to the stellar mass; in this regime, the halo structure is unconstrained because it is not contributing to the lensing, which is why the contours are very extended in the $\gamma(<R_e)$ direction. The shape of the degeneracy between $\gamma(<R_e)$ and $\alpha_c$ is also interesting, since it excludes the `vanilla' NFW+Chabrier IMF model for virtually all systems.}
\label{fig:contours}
\end{figure*}

\section{Discussion}
\label{sec:chap8sec4}

We have presented models of the dark and luminous mass structure of 12 massive ETG lenses, and found evidence that the sample can be characterised as having (a) haloes which deviate from dark-matter-only predictions and (b) heavy stellar-mass-to-light ratios. We now place our results in the context of previous work and consider their implications for the physics of ETG evolution.

\subsection{Stellar mass and the IMF}
\label{sec:chap8sec4sub2}

All of the EELs lenses have stellar masses which are heavier than those predicted from their photometry assuming a Milky-Way-like (i.e. Chabrier) IMF. This result is in agreement with a number of other studies of ETGs using strong lensing, dynamics and stellar population modelling (e.g. \citealp{Auger2010b,Cappellari2012,vanDokkum2010a}, but see also \citealp{Smith2013}), though we are able to make a stronger claim than previous strong lensing studies, as we have established that the need for a Salpeter-like IMF is not removed by the implementation of more flexible halo models than NFW profiles (though see also \citealp{Sonnenfeld2012}, in which a similar halo model was used for a single system, and similar conclusions drawn). We have furthermore shown that a quarter of the EELs lenses prefer to have radially declining stellar-mass-to-light ratio gradients, but that the presence of such gradients does not significantly reduce the mean IMF mismatch parameter of the population as a whole. This, then, underlines and strengthens previous indications that the IMFs of massive ETGs are heavier than those in Milky-Way-like systems -- and, in particular, that their IMFs may vary spatially \citep{MartinNavarro2015,vanDokkum2016} -- and that star formation must have occurred in correspondingly different physical conditions in these galaxies.

\subsection{Halo structure}

We find a majority of the EELs lenses to have haloes that are centrally steeper than the NFW expectation, with $\mu_{\gamma'} = 2.01_{-0.22}^{+0.19}$, whilst a small number of systems have virtually cored haloes, with $\mu_{\gamma'} = 0.10_{-0.10}^{+0.33}$. In the following subsections, we consider each of these groups in detail, and explore a possible explanation for the differences between them.

\subsubsection{Halo contraction in massive, isolated ETGs}

That seven out of the ten EELs with well-constrained dark matter properties have inner halo slopes that are significantly steeper than NFW suggests that these haloes may have been strongly contracted due to the inital cooling and infall of baryons \citep{Blumenthal1986,Gnedin2004}. Indeed, we use the adiabatic contraction model of \citet{Gnedin2004} to calculate the halo inner slope that would result from a scenario in which both dark and baryonic components begin as NFW profiles and the baryonic mass evolves into the inferred profile, and find that the typical halo inner slope that we would expect is $1.5 \lesssim \gamma \lesssim 2.0$, which is consistent with our inference on $\mu_{\gamma'}$ for this population. Our inferences are also in agreement with the more realistic study of \citet{Duffy2010}, which used high-resolution hydrodynamical simulations to measure the impact of adiabatic contraction on haloes in a cosmological context. That study found that the haloes of isolated galaxies become contracted such that $\gamma \sim 2$ in the presence of weak supernova feedback, with the action of either AGN or strong supernova feedback reducing this to $1.4 \leq \gamma \leq 2$. The EELs lenses are generally isolated systems (see Section 4.2.2); the cuspy structures that we have inferred may therefore be a result of these baryonic processes.

Previous studies of the halo structure of massive isolated ETGs are few, but are nevertheless consistent with our result. \citet{Sonnenfeld2012} used the larger-radius constraints available for the rare double source plane lens J0946+1006 to constrain the halo slope of that system to be $\gamma = 1.7 \pm 0.2$, which they interpreted as evidence for contraction due to the initial infall of gas. \citet{Grillo2012} combined Einstein radius and velocity dispersion measurements with simple stellar population models for a sample of 39 strong lenses to constrain the average logarithmic density slope of the population, which they found to be steep but dependent on the assumed IMF ($\gamma = 2 \pm 0.2$ for a Chabrier IMF and $1.7 \pm 0.5$ for a Salpeter IMF). Additionally, \citet{Napolitano2010} found an anticorrelation between the central dark matter density and galaxy size which they interpreted as evidence for halo cuspiness. In our modelling paradigm, which allows us to constrain the inner slopes of individual galaxies with higher precision, we find that the majority of the EELs are consistent with -- and strengthen -- these previous conclusions. In the following subsection, we now turn to the smaller number of systems whose haloes are not well described in this paradigm.

\subsubsection{Halo expansion in dense environments}

On the other hand, the evolution of the smaller group of systems with cored haloes (J0913, J1125 and J1323) appears to have been dominated by different processes from those explored in Section 4.2.1. Whilst the initial infall of gas onto the halo can cause contraction as described above, simulations have also shown that dynamical heating due to the accretion of satellite material and AGN-driven outflows can remove dark matter from the central regions of ETGs at later times \citep{Laporte2012,Martizzi2012}. It is possible that the systems with shallow haloes have experienced strong halo heating due to these processes; for instance, they may reside in denser environments and thus have experienced higher accretion rates, or have experienced more significant AGN activity. 

Firstly, we note that J1323 is unusual amongst the EELs as the lens is the bulge of a disky galaxy that has clear spiral structure at large radii and has an anomalously low central stellar velocity dispersion. It is therefore possible that this is a fundamentally different type of galaxy from the other EELs lenses, and may have been primarily subject to different physical processes in its central regions. The result of \citet{Dutton2013}, which showed that strong lensing and gas kinematics favour $\gamma < 1$ in six massive spiral galaxies, is consistent with this picture. 

We investigate the possibility that the shallow cusps in the haloes of the other two systems (J0913 and J1125) are due to particularly high rates of accretion of satellite material in dense environments by comparing their local environments with those of the other EELs. For each EEL, we select all galaxies within a physical projected radius of X = (0.5,1,1.5) Mpc from the lens and a redshift within 0.05 of the lens using the SDSS/DR9 \texttt{PhotoObjAll}, \texttt{Photoz} and \texttt{SpecPhotoAll} tables, where we verify the robustness of our results by considering  the three different radii and both photometric and spectroscopic redshift catalogues. 

Whilst J0913's environment is indistinguishable from those of the other EELs when this measure is used, J1125 is a striking outlier, with almost twice as many nearby galaxies as any other lens. Interestingly, J1606 has the next highest number of nearby galaxies, offering (weaker) evidence that this system may also occupy -- and indeed be the `central' of -- a dense environment; in our inference, we find the central dark matter fraction of this lens to be extremely low ($f_{DM} = 0.03 \pm 0.02$) and the halo inner slope unconstrained, which would be consistent with a scenario in which dark matter has been removed from the central regions of this system due to dynamical heating. In Figure 9, we show VI colour images of J1125, J1606 and J0913 (which represents a `typical' isolated EEL); these qualitatively support our initial impression that J1125 and J1606 may occupy denser environments than the other systems, and suggest some intriguing hints about the evolution of the dark haloes of ETGs.  A number of observations-based studies have also found the haloes of brightest-cluster galaxies (BCGs) to be shallower than NFW in their central regions \citep{Newman2013,Oldham2016b}, providing further tentative evidence for a role for the large-scale environment in shaping the halo (see also Section 4.2.3). 

\subsubsection{Systematically varying halo structure?}
\label{sec:chap8sec4sub1sub2}

The results of the previous two subsections motivate us to consider that ETG halo structure varies systematically with large-scale environment. As suggested by \citet{Newman2015}, ETGs in the centres of dense environments such as groups and clusters may have less cuspy haloes than their counterparts in the field due to the higher accretion and merging rates experienced by the former, leading to stronger dynamical heating. This is a scenario which could coherently explain previous results on the scale of isolated ETGs (\citealp{Grillo2012,Sonnenfeld2012,Sonnenfeld2015} and Section 4.2.1) and group- and cluster-scale ETGs (\citealp{Newman2015,Newman2013,Oldham2016b} and Section 4.2.2). Since the EELs for which we infer cuspy halo structures appear \emph{not} to reside in the centres of dense groups, whereas one of the two ETGs with sub-NFW haloes \emph{is} potentially associated with a group (we exclude J1323 from this discussion due to its clear morphological differences from the rest of the sample), this scenario is broadly consistent with our results. In Figure~\ref{fig:chap8fig6}, we put the EELs in the context of previous results  on cluster- and group-scales \citep{Newman2013,Oldham2016b,Newman2015,Sonnenfeld2012, Sonnenfeld2015} to show that tentative evidence for such an environmental dependence does seem to persist to higher masses. 

This sequence of increasing halo expansion in increasingly dense environments suggests a scenario in which the dark matter haloes of ETGs become initially contracted due to the infall of gas, and are subsequently heated during the accretion events which also cause them to grow in size, with the degree to which this heating erases the initial contraction signature dependent on the amount of accretion that a particular galaxy experiences. On the other hand, the fact that we do not have an explanation for the sub-NFW halo in J0913 highlights the possibility that other processes, such as AGN-driven outflows, may also play an important role in evolving the halo. Based on current results, the scenario suggested here is compelling, but more studies on all mass scales are needed to test it further. We plan to extend the models presented here to a $\sim 5$ times larger sample of field ETG lenses from SLACS in a future work. We also note the progress that is being made in constraining halo properties in galaxy clusters \citep{Jauzac2017}, which will be highly complementary to our work.

\subsubsection{Limitations at large radii}
\label{sec:chap8sec4sub1sub3}

In terms of the halo, we note that the main limitation of our current modelling paradigm is that we are unable to recover the halo structure at large radii due to the absence of mass probes on these scales. The distribution of $R_E/R_{e}$ for the EELs has a median of $\sim 0.5$, such that the lensing only probes the mass in the central regions; though the velocity dispersion is measured over a larger aperture ($\sim 1 R_e$), this is luminosity-weighted which means that it is also most constraining in the centre. Since the majority of the lenses are isolated systems, there are no substantial dynamical tracer populations such as satellite galaxies at larger radii, and the weak lensing signal or X-ray emission of an individual object at $z\sim 0.3$ would be prohibitively low to measure. We are therefore unable to make meaningful inference on the halo scale radius $r_s$, and find that its posterior distribution either resembles the prior (which is uniform for $0 \leqslant r_s \leqslant 200$kpc; this is the result for most systems), or becomes unreasonably small (and comparable to the effective radius of the light). The former scenario is uninformative and prohibits inference on larger-radius mass measures such as the virial mass; the latter scenario leads to virial masses which are unreasonably low, and may be a sign of a mismatch between the true halo structure and the gNFW model that we are using to describe it (indeed, we see a similar effect in simulations when true and model haloes become particularly mismatched; see the Appendix). 

We note that previous strong lensing studies have also generally predicted halo masses below abundance matching expectations based on stellar population modelling of the galaxy colours \citep{Auger2010b,Sonnenfeld2015}.  One option to make progress would be to use abundance matching expectations for the virial mass as a large-radius mass constraint; however, these relations have a large scatter and so do not have much constraining power. Moreover, the fact that we explicitly find stellar masses which require a heavier IMF than that assumed in the construction of abundance matching relations (i.e. heavier than a Chabrier IMF) makes the physical reasoning behind adding this constraint tenuous. We suggest that a better way to connect these high-precision measurements of the inner halo structure with constraints on larger scales will be statistically, by comparing inferences on samples of strong lenses such as the EELs with inferences on samples in which, for instance, the weak lensing signal is significant \citep{SonnenfeldInPrep}.  It is likely that making this connection will also require the investigation and adoption of more flexible halo models than the gNFW profile.

\subsubsection{Halo structure and star formation physics}

To facilitate comparison with previous observation- and simulation-based studies, we compute the projected dark matter fractions $f_{DM}$ within 5 kpc. Recently, \citet{Xu2016} compared the dark matter fractions of ETGs in the Illustris simulation with those inferred by \citet{Sonnenfeld2015} for the SLACS+SL2S strong lens samples, and found that the former have systematically higher central dark matter fractions at fixed velocity dispersion ($f_{DM} \sim 0.4 - 0.5$, compared to $f_{DM} \sim 0.2 - 0.3$ for the SLACS+SL2S samples at $\sigma_{\star} \sim 220$ kms$^{-1}$, the median velocity dispersion of the EELs). In that comparison, however, it was not possible to identify whether this tension was due to limitations in (a) the strong lens modelling or (b) the star formation and feedback models in the simulation. Here, our use of more extensive lensing information and more general mass models (i.e. a gNFW halo instead of NFW, and a S\'ersic stellar mass profile instead of de Vaucouleurs) allows us to test the first possibility. Table 2 includes the dark matter fractions implied by our lens models, and Figure 11 shows where these inferences place the EELs relative to SLACS+SL2S and the Illustris ETGs. We find that, though our sample size is small and our lenses span only a small range in velocity dispersion, they appear to show a similar offset from the simulated ETGs to that of the SLACS+SL2S samples. It is possible that the modelling techniques of both \citet{Sonnenfeld2015} and this study are biased in a similar way, although this is unlikely given the large differences between them  -- most importantly, we use the full pixel information, whereas the former used aperture mass measurements as in Section 3.5. This result therefore seems to confirm the finding of \citet{Xu2016} that the star formation and feedback prescriptions implemented in the Illustris simulation do not predict realistic central dark matter fractions in ETGs. We intend to carry out a more detailed comparison with the new results from IllustrisTNG in the future.


\begin{figure}
\centering
\includegraphics[trim=100 30 100 20,clip,width=0.4\textwidth]{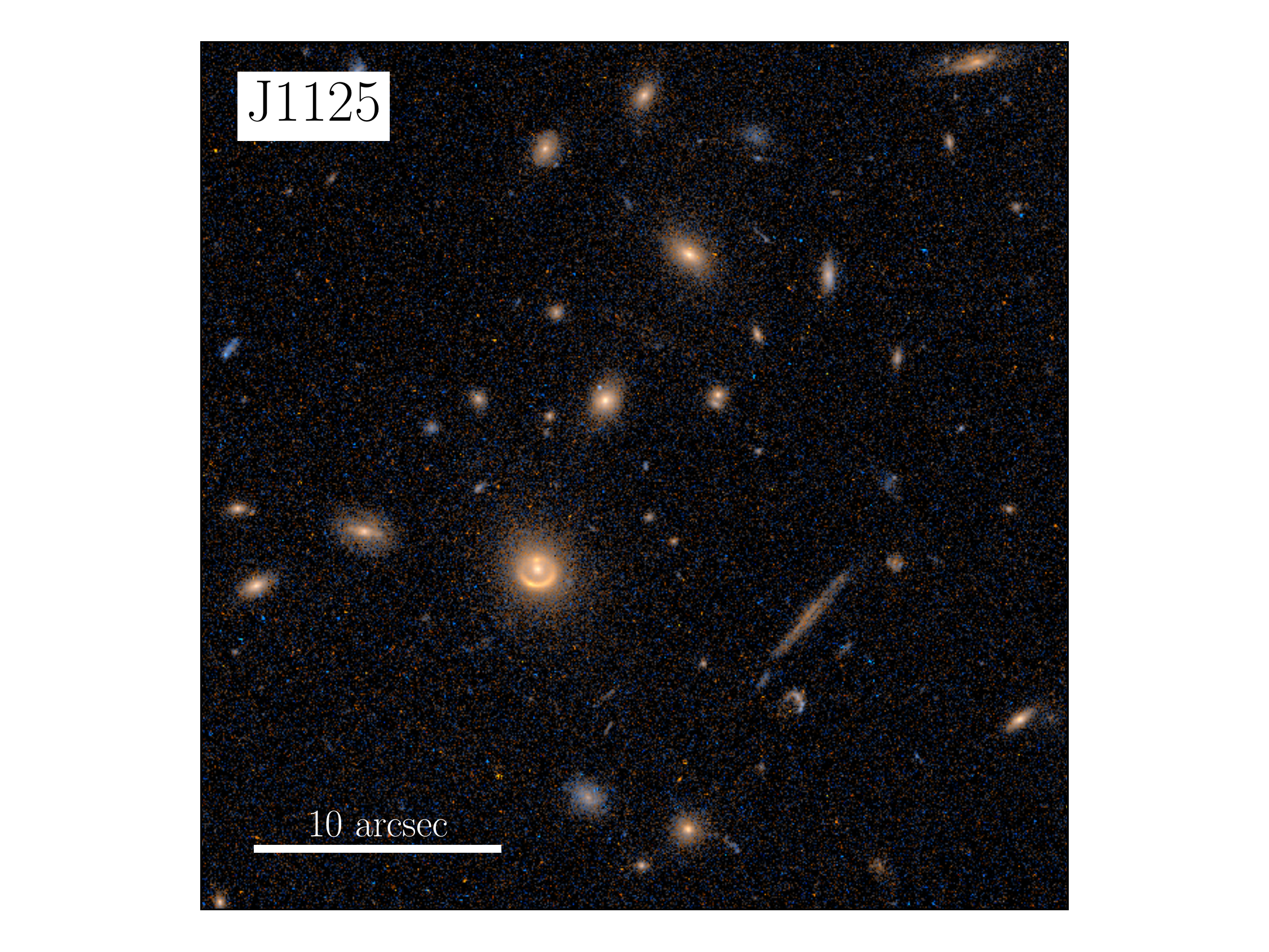}\hfill
\raisebox{0.1cm}{%
\includegraphics[trim=22 25 24 20,clip,width=0.4\textwidth]{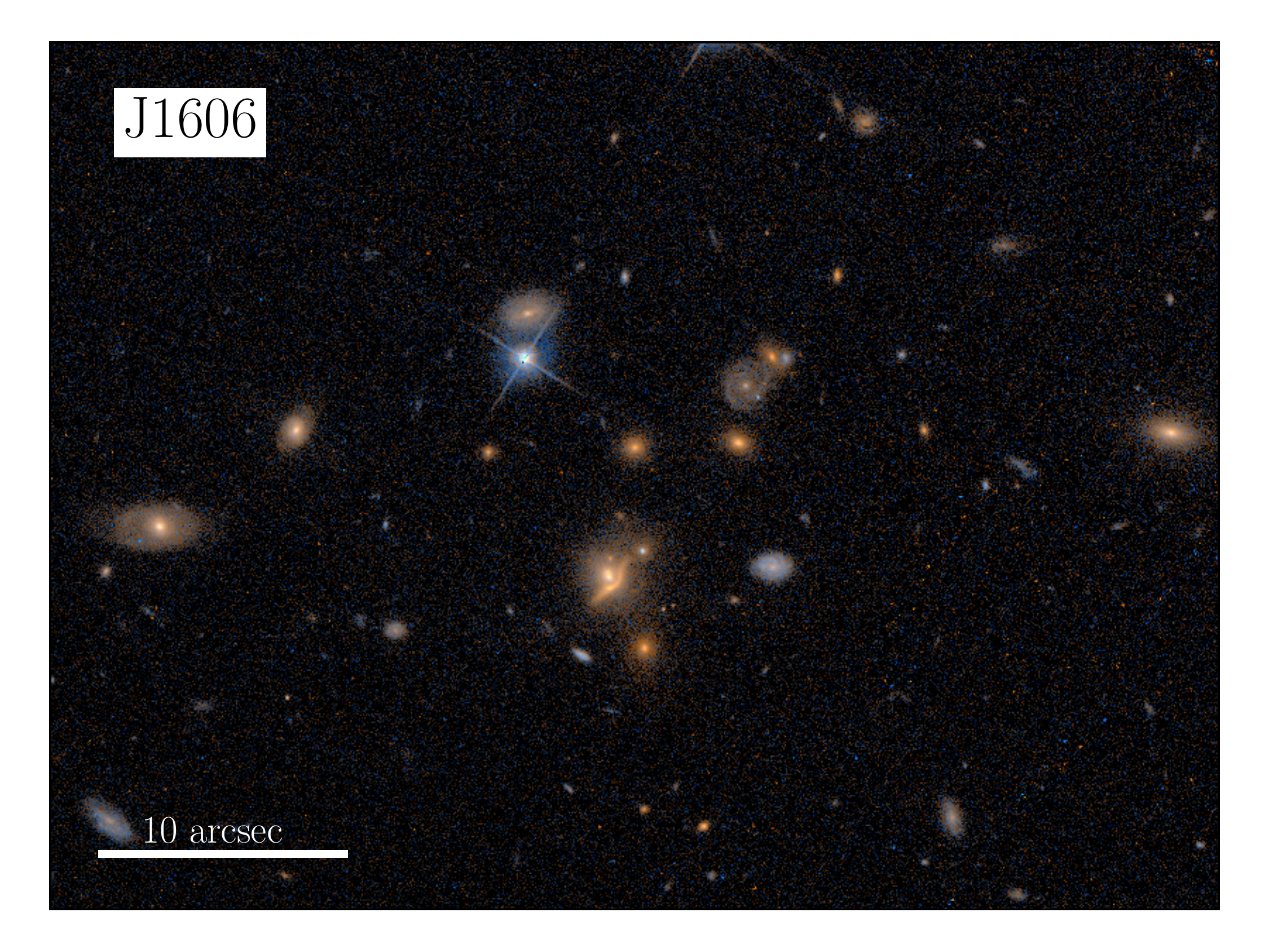}}\hfill
\raisebox{0.4cm}{
\includegraphics[trim=100 20 100 35,clip,width=0.4\textwidth]{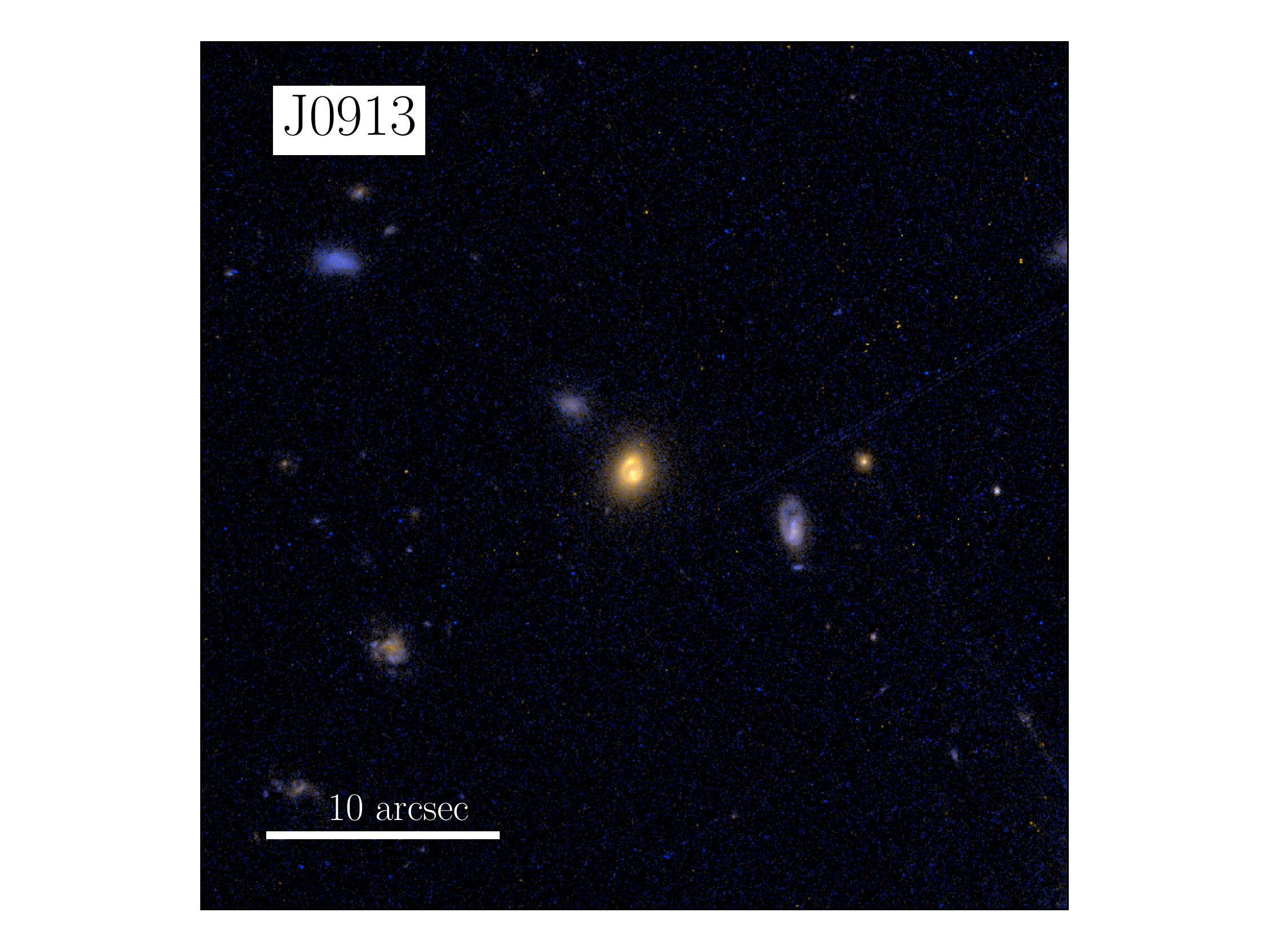}}\hfill
\caption{HST VI images showing the local environments of the EELs. J1125 (top) appears to occupy a denser environment than the other EELs, and this is supported by simple counts of the number of galaxies within a fixed projected distance and redshift range from the lens using SDSS. We suggest that the shallow central dark matter slope in this lens may be a result of its active accretion history as the massive member of a galaxy group. J1606 (middle) also appears to occupy a relatively dense environment, and has a very low central dark matter fraction and an unconstrained halo slope, whilst J0913 (bottom) has an environment typical of the remaining EELs (i.e. largely isolated). Our inferences are consistent with a scenario in which the central halo structure is related to the large-scale environment, though our small sample size means that we cannot make a strong claim.}
\label{fig:environments}
\end{figure}

\begin{figure}
\centering
\includegraphics[trim=20 20 20 20,clip,width=0.49\textwidth]{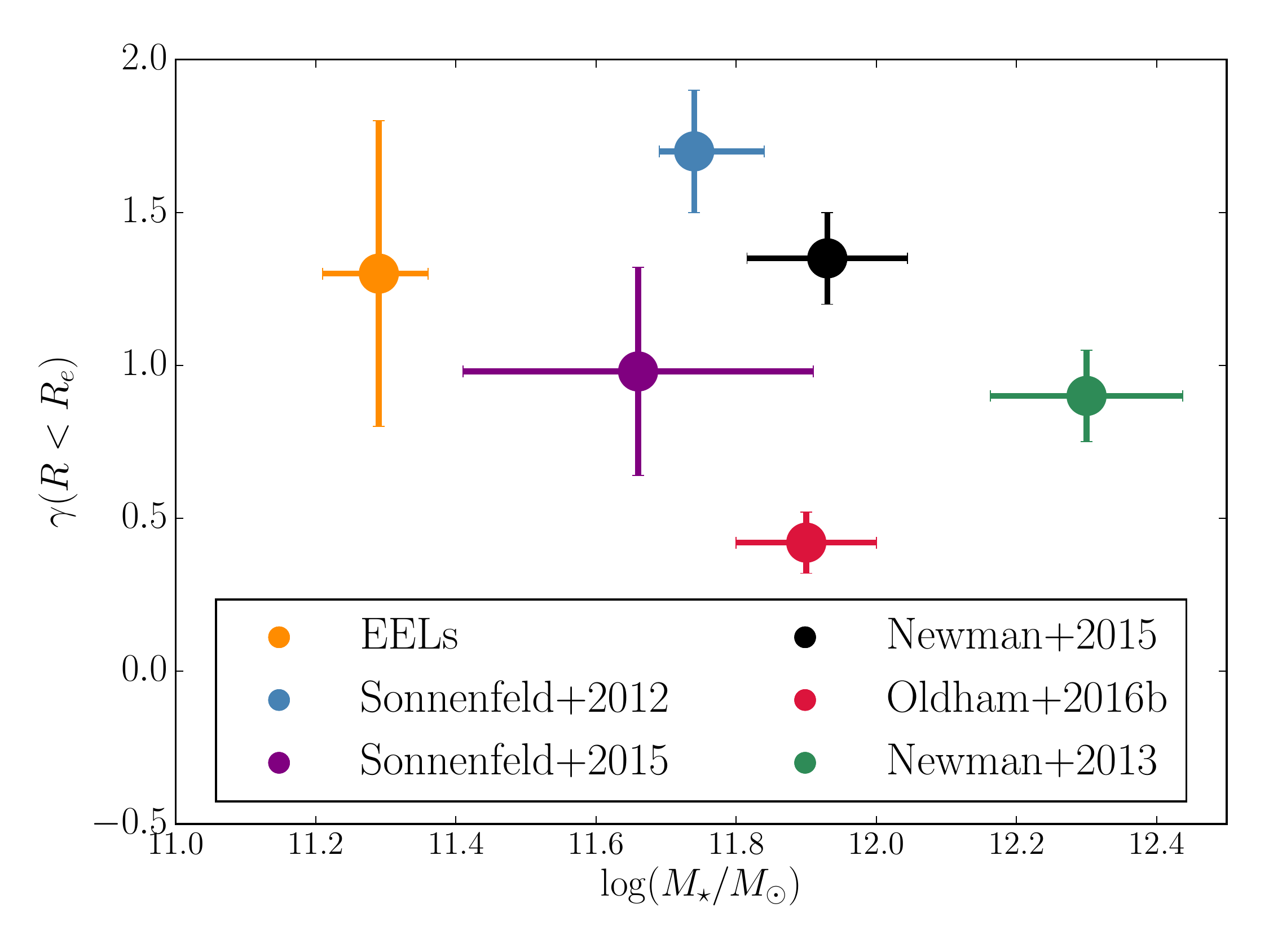}\hfill
\caption[Evidence for an environmentally-dependent halo structure]{A summary of inferences on the stellar mass and dark matter halo slope for ETGs. The mean mass-weighted halo slope within the effective radius appears to become increasingly shallow in ETGs in denser environments. This may reflect a real environmental dependence of the relative importance of different baryonic processes, such that the dark halo is more significantly heated in dense environments where the rate of merging and accretion events is higher.}
\label{fig:chap8fig6}
\end{figure}

\begin{figure}
\centering
\includegraphics[trim=10 10 10 10,clip,width=0.49\textwidth]{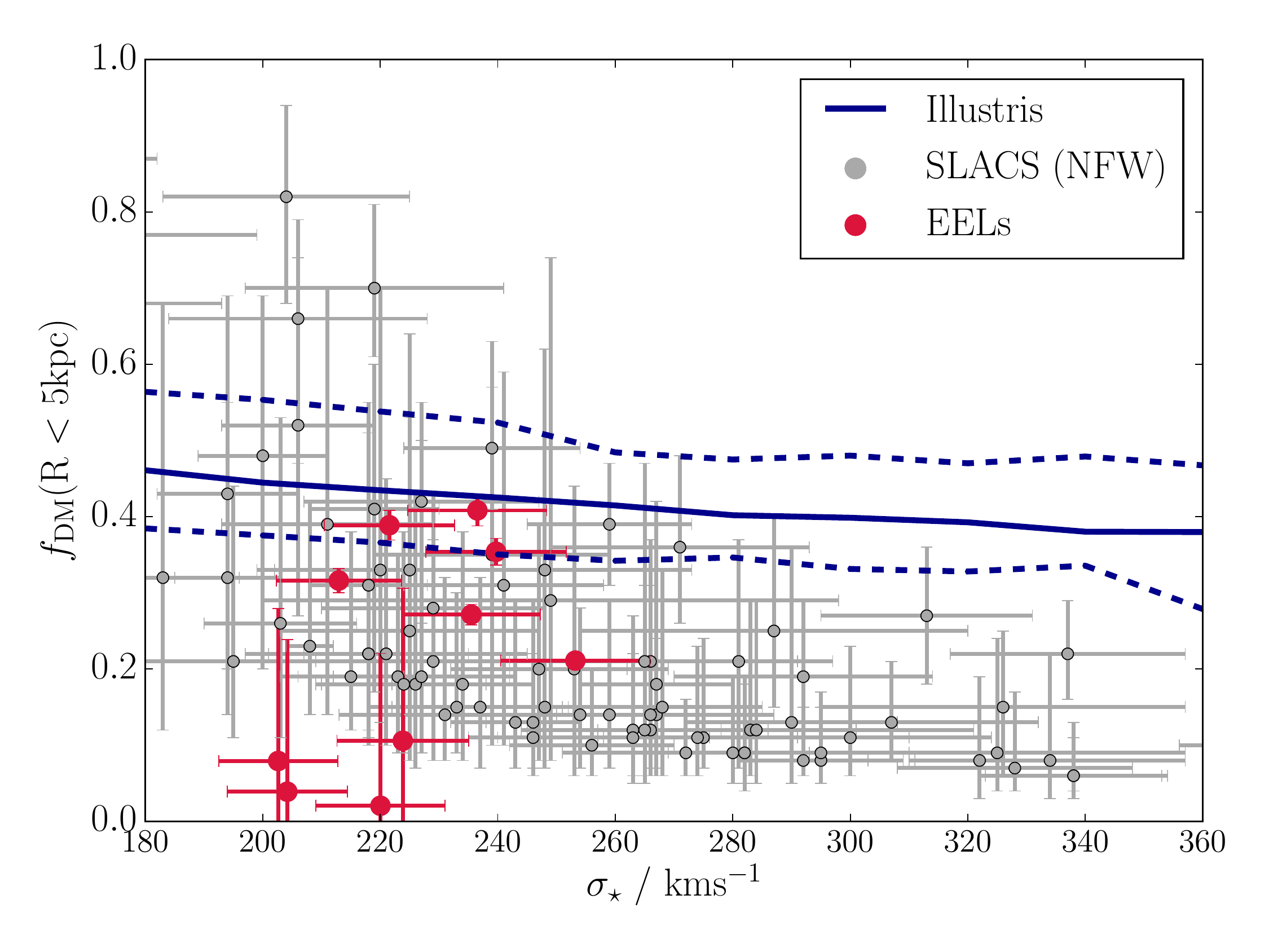}\hfill
\caption{Comparison of the projected dark matter fractions within 5 kpc, $f_{DM}(<5\mathrm{kpc})$, of the EELs and SLACS lenses \citep{Sonnenfeld2015} and ETGs in the Illustris simulation \citep{Xu2016}. Similarly to the SLACS lenses, the EELs lenses appear to have significantly lower central projected dark matter fractions than the simulated galaxies. Note that two lenses, J1323 and J1446, have velocity dispersions $\sigma_{\star} < 180 \mathrm{kms}^{-1}$ and so are not shown on this plot.}
\label{fig:hypers}
\end{figure}

\subsection{The evolution of massive ETGs}
\label{sec:chap8sec4sub3}

We have found evidence that massive ETGs have heavy stellar-mass-to-light ratios, that a significant fraction prefer radially declining stellar-mass-to-light ratios, and that there is a dispersion in inner halo slope which may correlate with environment. These results are consistent with a scenario in which the halo structure varies systematically across the ETG population and the stellar populations formed in different physical conditions from Milky-Way-like systems, as presented in the previous sections. We now attempt to draw these two results together to develop a coherent picture for the evolution of both the dark and the luminous mass.

Firstly, the steep central halo slopes of the majority of EELs suggest that these haloes have been contracted by the initial infall of gas. On the other hand, previous studies of ETGs in the centres of (a) groups and (b) clusters have found theirs haloes to be (a) only mildly contracted and (b) expanded, and a simple investigation of the environments of the EELs with expanded haloes suggests that at least some of them may also reside in denser environments. This seems to indicate that the large-scale environment of an ETG may be important in determining the halo structure in its inner regions such that, though all haloes initially experience some contraction, those in denser environments are subsequently subject to stronger heating processes due to satellite accretion, which undoes some of the initial contraction. This raises the question of whether the IMF mismatch should also exhibit an environmental dependence. That is, the higher rate of the accretion of lower-mass, Milky-Way-like satellites -- which formed their stars in Milky-Way-like conditions and so have Milky-Way-like IMFs -- in BCGs relative to field galaxies might also reduce the effective IMF of the final system \citep[e.g.][]{Sonnenfeld2017} in addition to expanding its halo. If the accreted matter is preferentially deposited at large radii, it may also lead BCGs to have IMF gradients as a function of radius. On the other hand, the much greater spatial extent of the halo relative to the stellar mass may make it more sensitive to the large-scale environment. The role of AGN outflows in modifying the dark and luminous mass structure also remains unclear. These are questions which we cannot answer on the basis of the small sample of lenses presented in this work -- though we note that a number of our systems have both cuspy haloes \emph{and} high stellar masses -- but which we hope to address in future studies.

\section{Summary and conclusions}
\label{sec:chap8sec5}

We have combined pixel-based strong lens modelling with Jeans dynamical modelling to construct models for the dark and luminous mass structures of 12 isolated ETGs at $z\sim 0.3$ and have reached the following conclusions.

\begin{enumerate}
 \item Most of the EELs have dark matter haloes which are centrally steeper than the NFW profile and are consistent with being drawn from a Gaussian distribution with mean $\gamma(R<R_e) = 2.01_{-0.22}^{+0.19}$. This is consistent with a scenario in which the evolution of their haloes has been dominated by an early contraction event due to the initial infall of baryons.
 \item A small number of systems have shallow central halo slopes, which may be associated with differences in their morphologies and environments. Moreover, a comparison of the EELs -- which are generally isolated systems -- with previous inferences on the halo slope in group- and cluster-scale ETGs presents tentative evidence that the halo structure may vary systematically with environment. In this scenario, haloes in denser environments would be shallower as a consequence of their more active accretion histories and consequent strong dynamical heating.
 \item The EELs have stellar masses which universally require IMFs which are heavier than Chabrier, and are consistent with being drawn from a Gaussian population with a mismatch parameter relative to a Chabrier IMF $\alpha_c = 1.80 \pm 0.14$ (i.e. close to Salpeter). This demonstrates that allowing flexibility in the halo structure does not remove the need for non-Chabrier IMFs in ETGs. Furthermore, a quarter of our sample prefer radially declining stellar-mass-to-light ratio gradients. This supports previous results implying that the physical conditions in which ETGs first form stars may be fundamentally different to those in Milky Way.
 \end{enumerate}

\noindent
Extending these methods to larger lens samples will allow a more thorough hierarchical analysis of the trends within the population of isolated ETGs and will be the subject of a future work.

\section{Acknowledgements}
We thank the anonymous referee for their helpful comments on the manuscript, and Dandan Xu and Ale Sonnenfeld for useful discussions. Both authors thank the Science and Technology Facilities Council (STFC) for financial support in the form of a studentship (LJO) and an Ernest Rutherford Fellowship (MWA).

This paper uses data based on observations made with the NASA/ESA Hubble Space Telescope, obtained at the Space Telescope Science Institute, which is operated by the Association of Universities for Research in Astronomy, Inc., under NASA contract NAS 5-26555. These observations are associated with programme GO 13661 (PI: Auger). This paper also includes data obtained at the W.M. Keck Observatory, which is operated as a scientific partnership among the California Institute of Technology, the University of California and the National Aeronautics and Space Administration. The Observatory was made possible by the generous financial support of the W.M. Keck Foundation. The authors wish to recognize and acknowledge the very significant cultural role and reverence that the summit of Mauna Kea has always had within the indigenous Hawaiian community.  We are most fortunate to have the opportunity to conduct observations from this mountain.

\appendix
\section{Systematic tests}
\label{sec:chap8sec3sub4}

In the tables and figures in this paper, we show the statistical uncertainties from our MCMC sampling; however, in addition to the alternative models that we have explored in Section 3.4, there are a number of further systematic uncertainties introduced by our modelling assumptions. In reality, the following effects, neglected in our analysis, will lead to larger uncertainties:
\begin{enumerate}
\item Non-zero uncertainty on the stellar light distribution
\item Non-zero central black hole mass
\item The possiblity that the dark halo is flattened and/or offset from the stellar mass
\item Deviations of the halo mass structure from the parametric forms used
\item Deviations of the source luminosity structure from the parametric forms used.
\end{enumerate}
In this section, we quantify these additional sources of uncertainty.

\begin{figure*}
\centering
\includegraphics[trim=20 20 20 20,clip,width=0.5\textwidth]{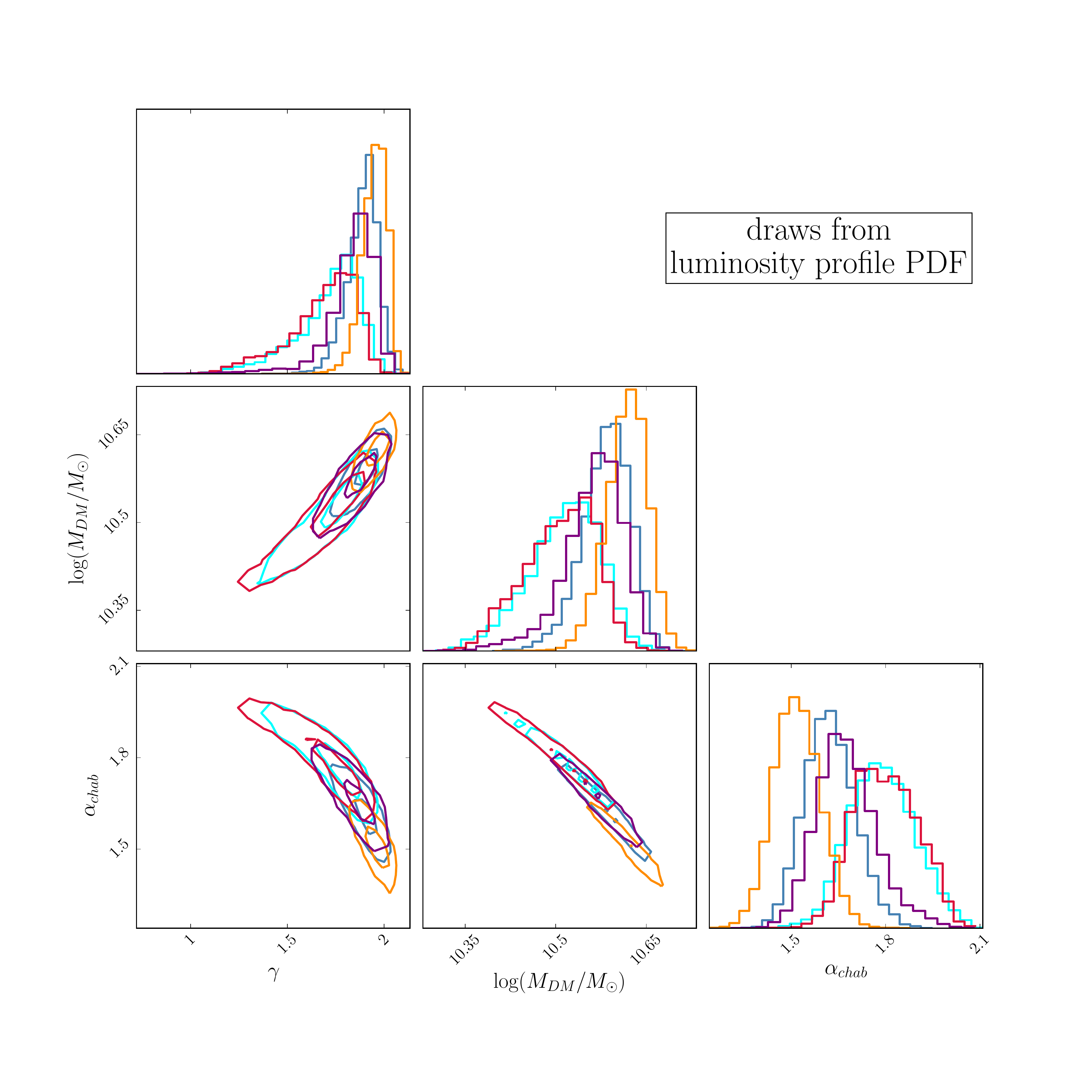}\hfill
\includegraphics[trim=20 20 20 20,clip,width=0.5\textwidth]{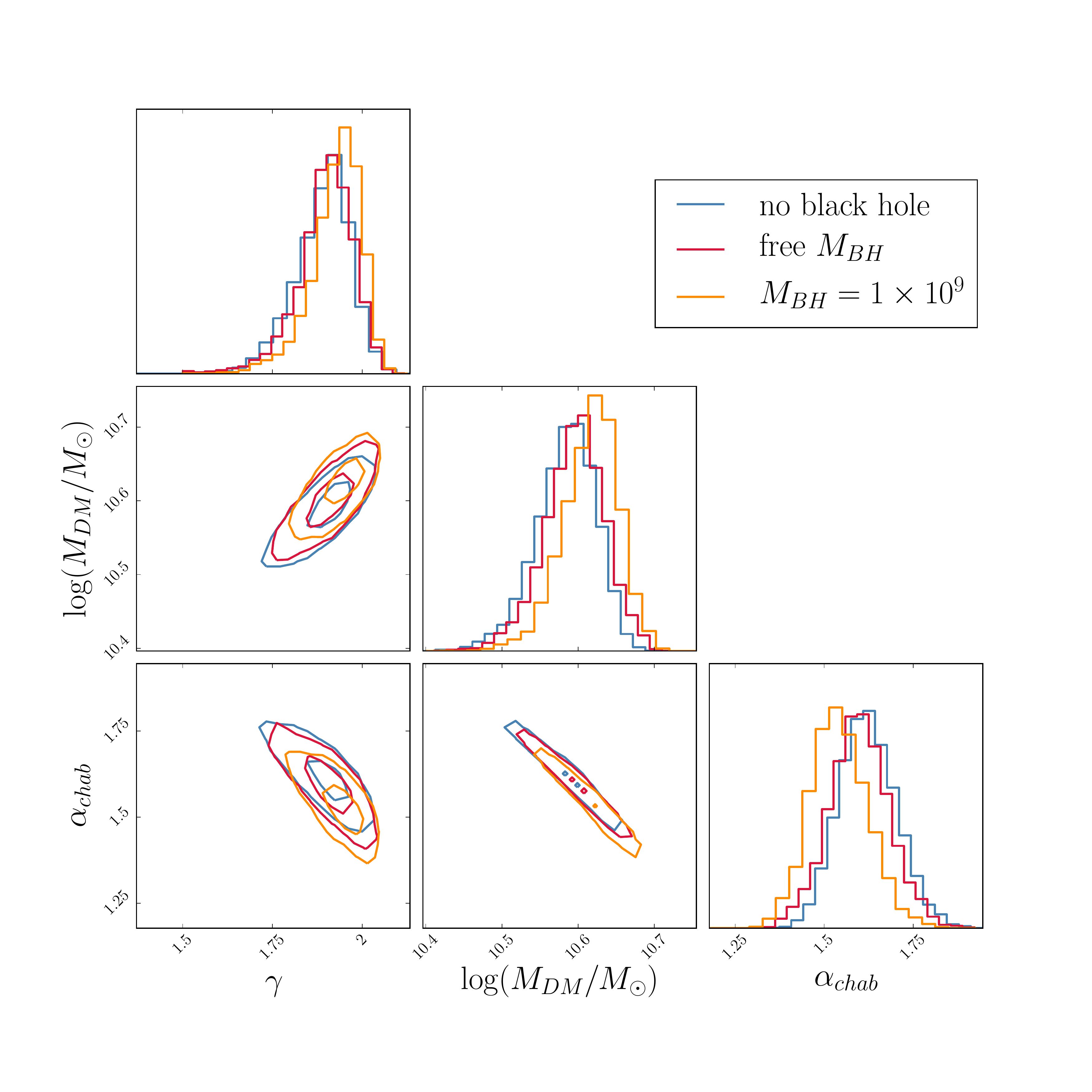}\hfill
\includegraphics[trim=20 20 20 20,clip,width=0.32\textwidth]{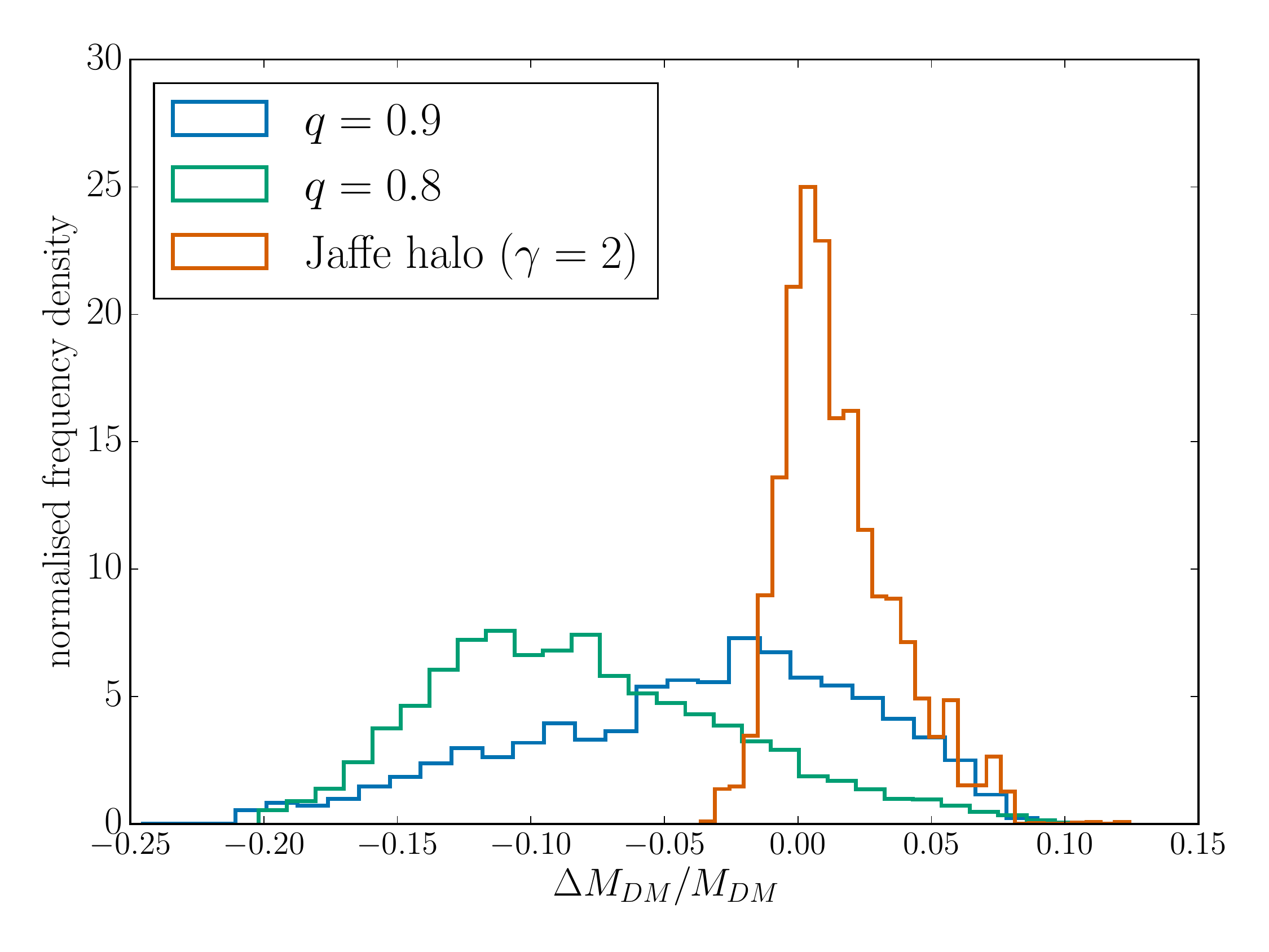}\hfill
\includegraphics[trim=20 20 20 20,clip,width=0.32\textwidth]{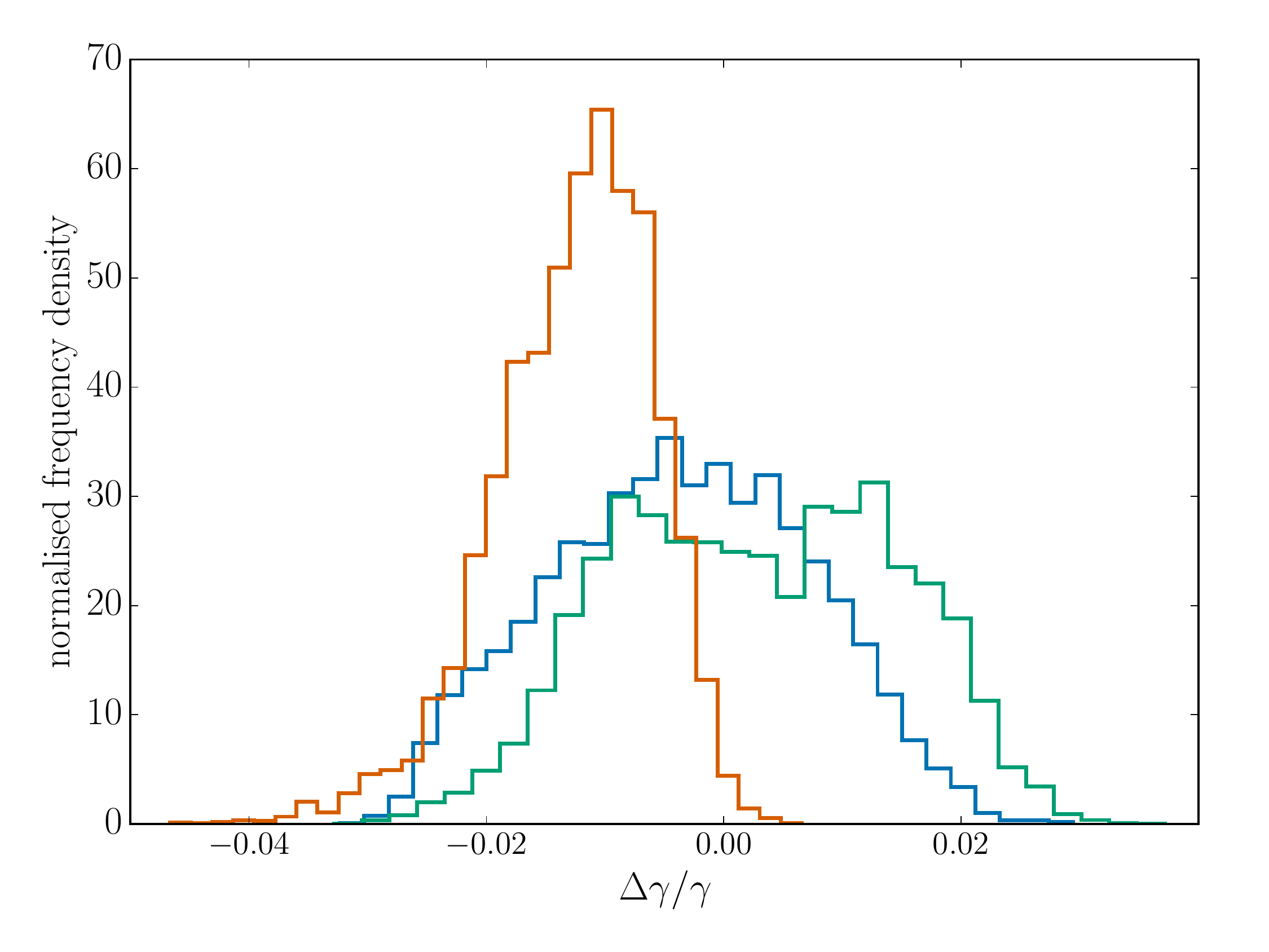}\hfill
\includegraphics[trim=20 20 20 20,clip,width=0.32\textwidth]{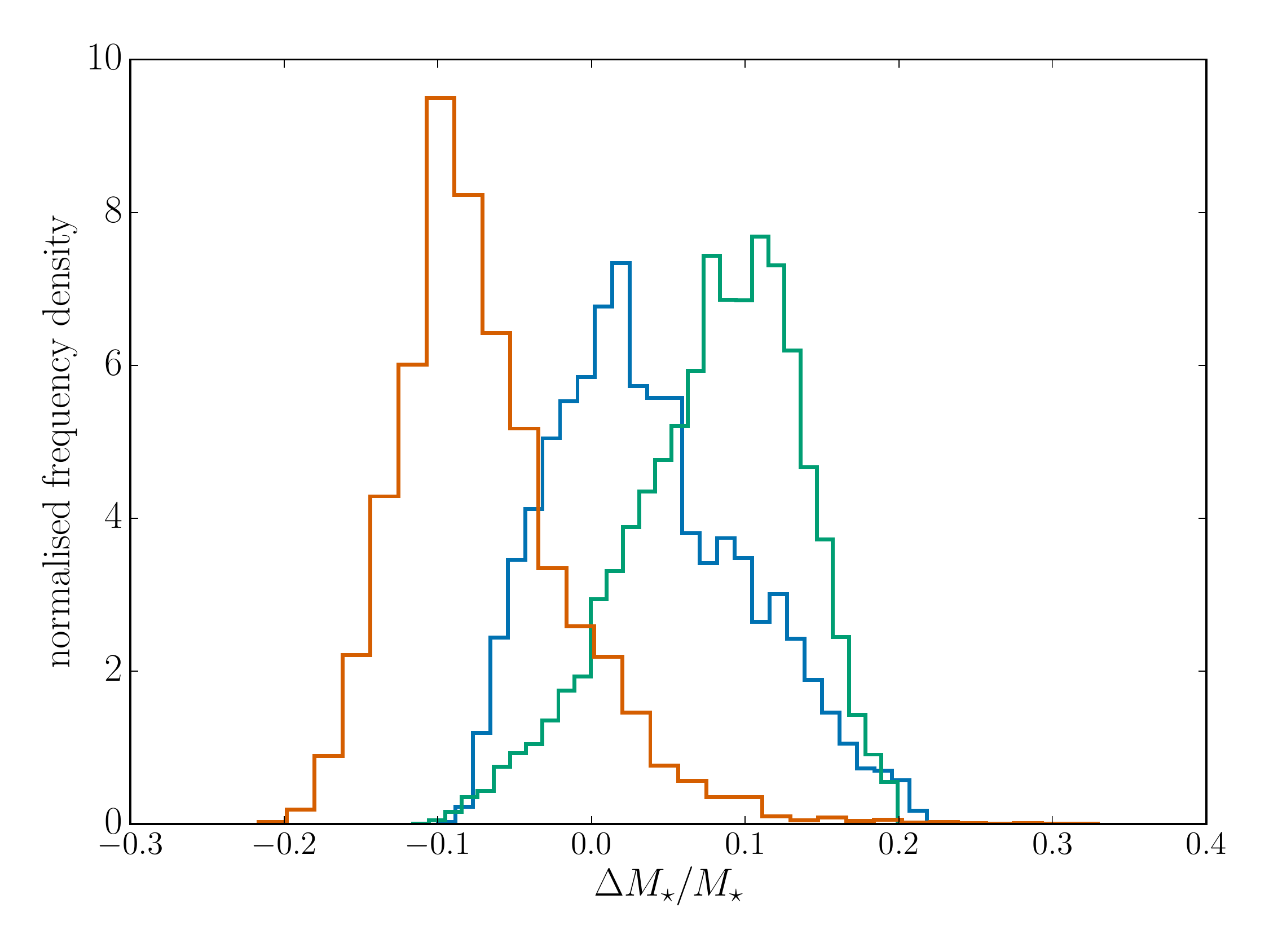}\hfill
\caption[Systematics in the construction of lens models for the EELs]{Systematics in the construction of lens models for the EELs. Drawing from the posterior distribution of the lens light parameters introduces additional uncertainty on the mass model parameters that are of the order to the statistical uncertainties (upper left); neglecting a massive central black hole has negligible impact on the inference (upper right). Assuming the halo is spherical introduces a small systematic bias into the inference (though assuming the halo is concentric with the stellar mass could introduce a larger bias; lower row).}
\label{fig:chap8fig5}
\end{figure*}

\subsection{Stellar mass}
\label{sec:chap8sec3sub4sub1}

Firstly, to make the lens modelling computationally feasible, we carry out our lensing and dynamical modelling with the stellar light distribution fixed according to the maximum-posterior profile that was inferred in \citet{Oldham2017a}. That study simultaneously modelled the light from the lens and the source such that the two components were robustly disentangled; nevertheless, the lens light distribution is only known as a posterior probability distribution with some finite width. By treating this distribution as a delta function, we are underestimating our uncertainty on the mass parameters. To quantify this extra uncertainty, we randomly draw lens light distributions from the posteriors of a subset of the lenses and rerun the inference. Figure~\ref{fig:chap8fig5} (top left) shows the scatter in our inference on the mass parameters for 5 independent model runs for a single lens (note that we performed 15 runs overall, but show a representative sample here for clarity): we find additional uncertainties of $\pm 0.10$ in $\alpha_c$, $\pm 0.09$ in $\gamma$ and $\pm 0.04$ in $\log M_{proj}$. These are generally of the order of the statistical uncertainties on those parameters (i.e. small), and are typical across the sample.

\subsection{Black holes}
\label{sec:chap8sec3sub4sub2}

A second potential source of bias in our inference on the mass of the dark and luminous components is the fact that we are ignoring the presence of the central black hole. Based on the $M_{BH}-\sigma_{\star}$ relation \citep{McConnell2013}, the median velocity dispersion $\langle \sigma_{\star} \rangle = 223$ kms$^{-1}$ of the EELs lenses corresponds to a black hole mass $M_{BH} \sim 3.9 \times 10^{8} M_{\odot}$. In contrast, the median Einstein mass of the EELs lenses is $\langle M_{Ein} \rangle = 1.1 \times 10^{11} M_{\odot}$; the black hole mass therefore makes a negligible ($\sim 0.3 \%$) contribution to the total lensing mass. Equally, since our inference on the central potential is driven by information at the Einstein radius, the deviation of the potential from our models at $R=0$ should not have a significant effect.

Nevertheless, given that we infer steep total mass slopes, it is important to quantify any potential bias that could arise from ignoring this contribution. We therefore run two additional tests for a subset of lenses. First, we run modified models in which a black hole is included. We assume the black hole to be concentric with the halo and baryonic components and infer its mass along with the other mass parameters; in this case, we are unable to obtain meaningful constraints on the black hole mass and find it to be consistent with zero. Given the very small contribution of any realistic black hole to the projected mass within the Einstein radius, this is not a surprising result. Secondly, we rerun a set of models which now include black holes with fixed mass. To check extreme cases, we consider black hole masses $M_{BH} = (1, 5, 10, 50) \times 10^{8} M_{\odot}$; in all these cases, our inference does not change within the statistical uncertainties. Figure~\ref{fig:chap8fig5} (top right) summarises these tests.

\subsection{Flattened and offset haloes}
\label{sec:chap8sec3sub4sub3}

A further possible source of bias is our assumption that the dark matter haloes are spherical and concentric with the light. It is possible that the halo may be flattened, though simulations suggest that the halo is generally rounder than the light \citep{Abadi2010,Zemp2012}. The median axis ratio for the EELs lens light profiles, $q = 0.8 \pm 0.1$, therefore sets a lower limit on the range of halo axis ratios that we might expect. The halo may also be offset from the stellar mass due to recent merger events: whilst CDM simulations constrain dark/light offsets to be less than the gravitational softening length of the simulations (350 pc; \citealp{Schaller2015}), offsets have been inferred to exist in some strong lensing galaxy clusters (\citealp{Harvey2017}, though see also \citealp{Massey2017}). 

Whilst we do not attempt to constrain the axis ratios or spatial offsets of the haloes in our sample -- we are unable to make meaningful inference on the axis ratio, and allowing spatial offsets would prohibit the use of simple dynamical models -- we construct a suite of simulated lenses to investigate the bias that is introduced by modelling flattened, offset haloes using spherical, non-offset lens models. We generate synthetic high-signal-to-noise HST-like images of fake lens systems in which the source light is given by a single S\'ersic profile and the lensing mass is the linear sum of (1) a stellar mass component with a spatially uniform stellar-mass-to-light ratio and a S\'ersic light profile and (2) a gNFW halo which is flattened and/or offset from the stellar mass. We consider halo axis ratios in the range $q_h = (0.8,1)$ to span the axis ratio range of the stellar light profiles of the EELs, and spatial offsets up to 0.1 kpc (corresponding to $\sim 2 \%$ of the half-light radius of a typical EEL). 

Figure~\ref{fig:chap8fig5} (bottom) shows the results of these tests. For the flattened, concentric case, we find that inference on $M_{\star}$, $\gamma$ and $\log M_{DM}$ is robust against the simplifying assumption of a spherical halo. For offset haloes, on the other hand, we find that the accuracy of our inference on the mass parameters depends on the cuspiness of the halo but is relatively insensitive to the flattening. Our recovery of the halo slope deteriorates from $\Delta \gamma / \gamma = 0.04$ for cuspy haloes ($\gamma = 2$) to $\Delta \gamma / \gamma = 0.25$ for NFW haloes ($\gamma = 1$). Our recovery of $\log M_{\star}$ and $\log M_{DM}$ is worse than this, however. For cuspy haloes ($\gamma = 2$), we
find $\Delta M_{\star} / M_{\star} \sim 0.3$ and $\Delta M_{DM} / M_{DM} \sim -0.3$, whereas for NFW haloes ($\gamma = 1$), we underestimate $M_{\star}$ and overestimate $M_{DM}$ by a similar fraction. If the haloes of the EELs lenses are significantly offset from the baryonic distributions, then, this could lead to a systematic bias in our results. We emphasise, however, that there is no theoretical or observational evidence for the existence of such offsets in field galaxies, and that the BCGs in which offsets have been tentatively inferred occupy very different environments from the EELs lenses, which, as isolated systems, are likely to be more dynamically relaxed than BCGs at cluster centres.

\subsection{Non-gNFW haloes}
\label{sec:chap8sec3sub4sub4}

We also use our simulations to test the bias introduced by modelling non-gNFW haloes using a gNFW profile. Whilst the NFW profile provides a good description of dark haloes in the absence of baryonic physics, simulations including prescriptions for baryonic processes which reproduce many of the observed scaling relations in galaxies on dwarf to Milky Way mass scales are better described by a yet more general form of the NFW profile in which the outer slope and break softening also vary (though the variation can be well parameterised by the stellar-to-halo mass ratio; \citealp{diCintio2014}). Comparable studies have not yet been carried out for massive ETGs (as it is much more computationally expensive to simulate such massive galaxies in large numbers), but it is unlikely that a gNFW profile can provide a complete description of the halo on these scales either. Furthermore, when we use gNFW profiles to fit the adiabatically contracted haloes that we calculated for the EELs (Section~\ref{sec:chap8sec3}), we find residuals of $\sim 20 \%$ at intermediate radii; this is because adiabatic contraction acts to `pinch' a pristine NFW halo such that the halo remains NFW-like in these regions, whereas the gNFW profile forces the halo slope to steepen monotonically with radius. 

We test the robustness of our interpretation of $\gamma$ as the inner slope of the dark matter profile by simulating lenses with haloes described by Jaffe profiles (with inner and outer slopes $\gamma = 2$ and $\gamma = 4$ respectively, and a softer break than our gNFW parameterisation). We find our inference on $\gamma$ using gNFW models recovers truth within $4\%$ (Figure~\ref{fig:chap8fig5}, bottom row). This indicates that $\gamma$ is indeed recovering the inner halo slope, regardless of the slope at larger radii and the strength of the break. Though lensing is sensitive to the \emph{projected} mass, the contribution of the outer regions, where the halo structure starts to differ substantially, is sufficiently small as to have a negligible effect on our inference on the properties of the central regions.

\subsection{Parametric versus pixellated models}

A further potential source of uncertainty is our requirement that the background galaxies follow S\'ersic light distributions. Whilst this is not an unreasonable assumption given the fact that these sources are also ETGs (i.e. they are passive), it does restrict the parameter space to some extent. We therefore construct pixellated source models for all the systems, using pixellated lens modelling techniques based closely on \citet{Vegetti2009}. The inference in this case is completely independent of the parametric models, and the source structure is subject to very different assumptions as it is now governed by a regularisation rather than an analytic profile. In general, we find that our inference on the mass structure scatters around the posteriors inferred in our parametric models with a scatter which is of the order of the statistical uncertainties; therefore, the choice of source model does not appear to be a significant source of uncertainty in our inference.

\end{document}